\documentclass[final,1p,times]{elsarticle}
\usepackage{graphicx}
\usepackage{amssymb}
\usepackage{lineno}

\begin{document}

\begin{frontmatter}

\title{Superinsulator-Superconductor Duality in Two Dimensions}

\author[USA,ISP]{Tatyana I. Baturina}
\ead{tatbat@isp.nsc.ru}
\author[USA]{Valerii M. Vinokur\corref{VMV}}%
\ead{vinokour@anl.gov}

\address[USA]{Materials Science Division, Argonne National Laboratory, Argonne,
Illinois 60439, USA}
\address[ISP]{A. V. Rzhanov Institute of Semiconductor Physics SB RAS,
13 Lavrentjev Avenue, Novosibirsk, 630090 Russia}
\cortext[VMV]{Corresponding Author}

\begin{abstract}
For nearly a half century the dominant orthodoxy has been that the only effect 
of the Cooper pairing is 
the state with zero resistivity at finite temperatures, \textit{superconductivity}.  
In this work we demonstrate that 
by the symmetry of the Heisenberg uncertainty principle relating the amplitude 
and phase of the superconducting order parameter, Cooper pairing can generate the dual state 
with zero conductivity in the finite temperature range, \textit{superinsulation}. 
We show that this duality realizes in the planar Josephson junction arrays (JJA)
via the duality between the Berezinskii-Kosterlitz-Thouless (BKT) transition 
in the vortex-antivortex plasma, resulting 
in phase-coherent superconductivity below the transition temperature, 
and the charge-BKT transition occurring in the
insulating state of JJA and marking formation 
of the low-temperature charge-BKT state, \textit{superinsulation}.
We find that in disordered superconducting films that are on the brink of
superconductor-insulator transition the Coulomb forces between 
the charges acquire two-dimensional character, i.e. the corresponding 
interaction energy depends logarithmically upon charge separation, 
bringing the same vortex-charge-BKT transition duality,
and realization of superinsulation in disordered films as the low-temperature charge-BKT state.  
Finally, we discuss 
possible applications and utilizations of  superconductivity-superinsulation duality.
\end{abstract}


\end{frontmatter}

\section{Introduction}

Hundred years ago the discovery of superconductivity,
the state possessing a zero resistance at low but finite temperatures,
opened a new era in physics.
The discovery crowned a systematic study of electronic transport
at lowest temperatures available at the time undertaken
at Leiden University by Professor Kamerlingh Onnes, who strived to uncover microscopic
mechanisms of conductivity in metals.
It was recognized since 1905 that conductivity can be described
by the Drude formula  $\sigma=ne^2\tau/m$, where $n$ is the charge carrier density,
$-e$ is the electron charge, $\tau$ is the scattering time, and $m$
is the mass of the charge carrier.
A natural hypothesis was that in pure metals the scattering time will start to grow indefinitely
upon approaching the absolute zero temperature,
resulting in infinite conductivity.
However Lord Kelvin argued that mobile charges can freeze out near
the zero temperature so that the charge density $n$ in the Drude formula
goes to zero and that this effect may win over the increase of  $\tau$.
Kamerlingh Onnes' finding ruled out Lord Kelvin hypothesis, and,
one can say -- adopting extremely loose and non-rigorous interpretation
of known now microscopic mechanisms of superconductivity -- that superconductivity
indeed complies with the $\tau$-divergence notion. It is essential for the whole
concept of superconductivity that $\tau$ becomes infinite at \textit{finite} temperatures.  
The idea about the mobile charges gradually freezing out as temperature tends to zero 
was found to realize in the resistivity of semiconductors (i.e. band insulators).  
The latter exhibit the metallic type of behavior at high temperatures but 
at low temperatures their charge density $n$ exponentially decreases
with the decreasing temperature and wins over the gradually
growing $\tau$.
The next step towards what might sound as a realization of the Lord Kelvin idea,
was the finding that there exist two-dimensional superconducting systems, 
namely, Josephson junction arrays, granular 
and homogeneously disordered films, which unite 
both types of behavior, superconducting and insulating.
The idea that the material with the Cooper pairing can be transformed into an insulator
traces back to Anderson~\cite{Anderson64}, who considered small 
superconductors coupled by Josephson link,
and was further discussed by Abeles~\cite{Abeles77} in the context of granular systems.
The analytical proof for the existence of superconductor-insulator transition (SIT) in 
granular superconductors was given by Efetov~\cite{Efetov1980}, 
and the first theory of the disorder-induced
superconductor-insulator transition in the 3D disordered Bose condensate 
was developed by Gold~\cite{Gold1983ZPhys,Gold1986PRA}.  
Ever since the study of the SIT, which is an exemplary quantum phase transition,
 has been enjoying an intense experimental and theoretical attention 
 and has become one of the mainstreams
 of contemporary condensed matter physics.
In this work we will describe the further advance in understanding the SIT,
the realization that at the transition point superconductivity transforms into 
a mirror image of superconductivity, the \textit{superinsulating state}, 
possessing \textit{infinite} resistivity at \textit{finite} temperatures.

The paper is organized as follows.  
Next Section establishes the possibility of superinsulation from the uncertainty principle.
In the Section 3 we discuss general features of two-dimensional
systems enabling experimental realization of the superinsulating state;  
in the Section 4 we briefly remind the 
properties of Berezinskii-Kosterlitz-Thouless (BKT) transition and the relation between 
this transition and formation of the superinsulating state, 
 presenting the phase diagram of the system in the 
close vicinity of superconductor-superinsulator transition;
the Section 5 is devoted to realization of 2D electrostatics in disordered superconducting 
films that brings to the existence the charge-BKT transition; 
in the Section 6 we discuss a possible microscopic mechanism behind the superconducting state; 
in the Section 7 we present a qualitative consideration of the conductivity in the
superinsulating state; 
Section 8 discusses possible applications and utilizations of the 
superconductor-superinsulator duality; 
and finally, we summarize the results of our work in the Section 9.

\section{Superconductivity and superinsulation from the uncertainty principle}

The possibility of the existence of the superinsulating state can be established from
the most fundamental quantum-mechanical standpoint.
Superconducting state is characterised by an order parameter, $\Psi=|\Psi|\exp(i\varphi)$, 
where  the phase, $\varphi$, is well defined across the whole system.
The uncertainties in the phase $\Delta\varphi$ and the number of Cooper pairs 
in the condensate, $N_{\mathrm{s}}=2|\Psi|^2 $, 
compete according to the Heisenberg uncertainty principle 
$\Delta\varphi \Delta N\geq 1$~\cite{Anderson64,Feynman,Sugahara85,Mooij2006}.
This brings about the basic property of a superconductor, 
its ability of carrying a loss-free current.
Indeed, the absence of an uncertainty in the phase,  $\Delta\varphi = 0$,
i.e. the definite phase, implies that the number of condensate particles is undefined.
It means that the particles that comprise condensate cannot scatter and dissipate energy,
since every scattering event can be viewed as the measurement,
and, in principle, could be used to determine the number of particles.
In two-dimensional systems, the global phase coherence leading to true 
superconductivity establishes itself at some finite temperature, $T_{\mathrm{SC}}$, 
which is below the temperature $T_{\mathrm{c}}$, 
where the finite modulus of the order parameter appears~\cite{Beasley1979,HalperinNelson1979}.
The phase coherence then remains stable against moderate perturbations: 
the loss-free current maintains as long as it does not exceed the
certain value $I_\mathrm{c}$ determined by the material parameters.

Now let us assume that we have managed to tune the parameters of a 
2D superconductor in such a manner that the Cooper pairs got pinned within the system.  
This would eliminate the uncertainty of the total charge setting $\Delta N = 0$.
The Heisenberg principle requires then the finite uncertainty of the global phase, 
i.e. pinning the Cooper pairs results in $\Delta\varphi\neq 0$.
This implies that due to the Josephson relation, $d \varphi /dt=(2e/\hbar)V$,
where $\hbar$ is the Planck constant, there may be a finite
voltage, $V$, across the whole system while the flowing of the current is blocked.
This establishes the possibility for the existence of a distinct superinsulating 
state~\cite{FVB,VinNature}, which is dual to the superconducting one and 
has zero conductivity over
the finite temperature range from the critical temperature $T_{\mathrm{SI}}$ down to $T=0$.
Of course, one has to bear in mind that the term ``zero conductivity"
is used in the same idealized sense as the ``zero resistance" of a two-dimensional superconductor.
We would like to emphasize here once again the conceptual 
difference between the notions of  the``superinsulator"
and merely the ``insulator,"  which is that the former has \textit{zero conductivity} 
in the finite temperatures range, while the latter 
would have possessed the finite, although exponentially small, linear conductivity 
$\sigma \propto \exp(-E_g/k_BT)$ ($E_g$ is the energy gap) at all finite
temperatures, except for $T=0$.

\section{Two-dimensional superconducting systems}

To gain an insight into the origin of a superinsulating state we will focus 
on two-dimensional tunable superconducting systems, 
Josephson junction arrays (JJA), 
the systems comprised of small superconducting islands connected by Josephson junctions.
A theoretical and experimental study of large regular JJAs 
has been one of the main directions of the condensed matter physics 
for decades~\cite{Minnhagen,Mooij1989,Mooij1990,FazioSchon1991,WeesDual,Mooij1992,
Tighe,Haviland1,KandaChargeSoliton,KandaCBKT,KandaSelfcapacitance,Mooij1996,Japan1,Japan2},
see~\cite{Review1,2DJJA_Review,FazioZant2001} for a review. 
The interest was motivated by the appeal of dealing with the experimentally 
accessible systems whose properties can be easily controlled by tuning two major parameters
quantifying the behavior of the array,
the Josephson coupling energy between the two adjacent superconducting islands comprising a JJA, 
$E_\mathrm{J}$, and the charging energy of a single junction, i.e. the Coulomb energy cost 
for transferring the Cooper pair between the neighbouring islands,
 $E_\mathrm{C} = (2e)^2/(2C)$, where $C$ is the junction capacitance~\cite{Anderson64}.
Moreover, JJA  offers a generic model that captures essential features
of two-dimensional disordered superconductors including homogeneously disordered 
and granular films.
The remarkable feature of JJAs is that they experience a phase transition, 
which historically was described in
terms of the zero-temperature transition between the superconducting and \textit{insulating} states, 
superconductor-to-insulator transition (SIT), which occurs as $E_{\mathrm C}$ 
compares to  $E_{\mathrm J}$.
In the arrays that are already near the critical region where 
$E_{\mathrm C}\approx E_{\mathrm J}$, the SIT can be also  induced
by the magnetic field.  
The SIT studies in JJAs were paralleled by the 
investigations on the superconductor-insulator transition in thin granular films, which
revealed the similar 
behavior~\cite{Abeles77,Dynes78,Efetov1980,White,Orr86,SugaharaLT18,Sugahara26L,
Widom,Jaeger,Barber90,Wu94,GoldmanGranular,Frydman,Barber06}. 
What more, even homogeneously disordered superconducting
films~\cite{Strongin70,HavGold,Hebard,ShaharOvadyahu,LiuGoldman,
Zvi1994,VFGIns,Marcovic98,VFG1998,Valles1999,VFGScaling,VFGPar2000,
ButkoAdams,Be1,BeInsPMR,BeBSIT,Shahar-act,TBJETPL,Hadacek,
TiNPhysB2005,SITInOKapitulnik,Shahar-Coll,Xiong2006,TiNQM,
TiNSIT,TiNPhysC,Zvi2008,TiNHA,Goldman2010,ChargeBKT,
Goldman2011,ShaharAngular,ShaharSteps,KalokLT26}  
and layered high-$T_c$ 
superconductors~\cite{Mandrus1991,Rosenbaum1992,Tanda1992,Beschoten1996,
LavrovSIT,VFG_JETPL2003,suFET}
exhibited all the wealth of the SIT-related phenomena which were viewed 
as characteristic to granular superconducting systems.
This brought about the idea that in the vicinity of SIT disorder 
may cause the electronic phase separation (often referred to as
the ``self-induced granularity") that realizes in a form 
the strong disorder induces an
inhomogeneous spatial structure of isolated superconducting islands
in thin homogeneously disordered 
films~\cite{Zvi1994,VFGIns,LarkinOvchin71n1,IoffeLarkin,MaLee,Imry,Feig2005}.
Numerical simulations of the homogeneously disordered
superconducting films confirmed that indeed in the presence of sufficiently strong disorder, 
the system breaks up into superconducting islands separated
by an insulating sea~\cite{GhosalPRL,GhosalPRB,DubiNat,Trivedi2011}.  
Recent scanning tunneling spectroscopy measurements of the local density of
states in TiN and InO films~\cite{BaturinaSTM08,InOSTM_NP} 
and in high-$T_c$ superconductors~\cite{KapitulnikSTM,YazdaniSTM}
offered strong support to this hypothesis.  We will discuss these issues in more detail in Section 1.4.

All the above studies showed that the Josephson junction arrays
indeed offer a generic model that captures most essential features
of the superconductor-insulator transition in a wide class of
systems ranging from artificially manufactured Josephson junction
arrays to superconducting granular systems and even the
homogeneously disordered superconducting films and allows for consideration of all of them
on the common ground.  The properties of the films were discussed in terms of the dimensionless 
[measured in the quantum units $e^2/(2\pi\hbar)$] conductance $g$, 
which played the role of the tuning parameter 
for the films replacing the ratio $E_{\mathrm J}/E_{\mathrm C}$ characterizing JJAs.
Large $g$ correspond to superconducting domain, at small $g$ films turn insulating.
And at some critical value $g_{\mathrm c}\simeq 1$ disordered films is viewed 
to undergo superconductor-insulator transition. 

Using JJA as an exemplary system allows to understand better the microscopic mechanism of the 
realization of the charge-phase duality of the uncertainty principle giving rise to superinsulation.
As had already been indicated in~\cite{FazioNatNV}  the charge-vortex duality 
reflects the duality between the 
Aharonov-Bohm~\cite{ArBohm} and Aharonov-Casher~\cite{ArCash} effects.
In the Aharonov-Bohm effect, the charges moving in a field-free region 
surrounding a magnetic flux acquires
the phase proportional to the number of flux quanta piercing 
the area. Accordingly, the Aharonov-Casher effect is the reverse~\cite{RezAr}:
magnetic vortices moving in a 2D system around a charge acquires a phase proportional to that charge.
Now, considering an insulating side, $E_{\mathrm C}\gg E_{\mathrm J}$, 
where all the Cooper pairs are pinned 
at the granules by Coulomb blockade, we can conjecture,
following the line of reasoning of~\cite{Ivanov2001,Averin2002}, that in the fluxons
tunnel \textit{coherently} through the Josephson links between the granules that corresponds to 
\textit{constructive} interference of their tunnelling paths i.e. to the process having 
the largest quantum mechanical amplitude.
Thus the phase synchronizes across the whole system remaining undefined.
In other words, one can possible view the superinsulating state 
where the current are completely blocked as
a state mediated by the coherent quantum phase slips discussed recently 
in the context of Josephson junction 
chains~\cite{Matveev2002,Pop2010} and disordered wire ring~\cite{Astafiev2012}. 
The uncertainty of the phase implies the superfluid state of fluxons at the insulating 
side dual to superfluid state 
of Cooper pair condensate, by the same token as uncertainty in the charge 
implies superconductivity as discussed above.
Indeed the possibility of measuring the relative phase at different junctions that lifts the phase uncertainty,
implies the destruction of their coherent motion, 
which in its turn would have meant that Cooper pairs are not blocked 
at their respective granules any more.

We would like to emphasize here the crucial role played by charge pinning 
in the formation of the superinsulating state. 
It has been already recognized earlier~\cite{Doniach1998} that at very low temperatures 
the system of vortices transforms 
into a vortex-superfluid state in which an infenitizimal current induces an infinite voltage, i.e. the system 
acquires an infinite resistance and should be called a superinsulator. 
However the fact that in the absence of the charge pinning
the charge fluctuations would destroy the superinsulating state was 
not appreciated in~\cite{Doniach1998}. 
 
\section{Superinsulation and Berezinskii-Kosterlitz-Thouless transition}

As early as in 1963 Salzberg and Prager~\cite{Salzberg1963},
in the course of their study of thermodynamics of the 2D electrolyte, 
derived the equation of state and noticed that due to logarithmic interaction 
between the `plus' and `minus' ions, there appears a singular temperature 
where the pressure of ions turns zero.  
They ruled that at this temperature the ion pair formation occurs.  
In 1970 Berezinskii~\cite{Berezinskii1970,Berezinskii1971} and later Kosterlitz 
and Thouless~\cite{KT1972,KT1973} offered more refined consideration
to what had become known ever since as the binding-unbinding 
Berezinskii-Kosterlitz-Thouless (BKT) phase transition.
The physical idea behind the transition is as follows~\cite{KT1972}.  
The particle-antiparticle attraction contributes the term 
of ${\cal E}_0\ln(r/r_0)$ into the free energy of the system, 
where ${\cal E}_0$ is the energy parameter characterizing the interaction 
and $r_0$ is the microscopic spatial cut off parameter.  
At the same time the entropy contribution to the free energy is 
$-k_{\mathrm{\scriptscriptstyle{B}}}T\ln[(r/r_0)^2]$, as there are $\simeq(r/r_0)^2$
ways of placing two particles within the distance $r$ apart.  
At low temperatures the attraction between the particles and antiparticles wins 
and they remain bounded.  
At $T_{\mathrm{\scriptscriptstyle{BKT}}}\simeq{\cal E}_0/(2k_{\mathrm{\scriptscriptstyle{B}}})$ 
the entropy term balances the attraction, 
and at $T>T_{\mathrm{\scriptscriptstyle{BKT}}}$ the particles and antiparticles get unbound.  
We thus see that the BKT transition is the consequence of the logarithmic interaction 
between the constitutive entities in two-dimensional systems 
(2D vortex-antivortex system, 2D charge-anticharge plasma, 
dislocation-antidislocation array),  
irrespectively to their particular nature.
Below we will demonstrate that it is the BKT transition that provides the mechanism by which the 2D JJA 
falls into either superconducting or superinsulating states at low temperatures.
The choice of the particular ground state is dictated by the relation between 
 $E_\mathrm{J}$ and $E_\mathrm{C}$~\cite{Abeles77,Efetov1980,Mooij1989,FazioSchon1991,
WeesDual,2DJJA_Review,FazioZant2001}.

The absence of phase coherence at high temperatures and its appearance 
at low temperatures in a strong coupling regime, $E_\mathrm{J} > E_\mathrm{C}$,
can be described in terms of vortex-antivortex plasma dynamics 
and the corresponding BKT transition.
In two-dimensional superconductors and Josephson junction arrays finite temperatures 
$T$ generate  fluctuational vortices and antivortices, the
number of former and the latter remaining equal in order to conserve ``flux neutrality."
At high temperatures, $T>T_{\scriptscriptstyle{\mathrm{BKT}}}$ vortices 
and antivortices diffuse freely, and since each vortex (antivortex) carries the phase $2\pi$
their motion gives rise to breaking down the global phase coherence. 
Below the BKT transition temperature vortices and antivortices get bound 
into the neutral dipole pairs and cannot diffuse independently any more.
As the phase gain when encircling a vortex-antivortex dipole is zero, 
the diffusion of the dipoles as a whole
does not cause phase fluctuations on large spatial scales, and the global phase coherence sets.
The crucial feature that ensures binding of vortex-antivortex pairs 
and setting down the phase coherence
at low temperatures is the so-called vortex-antivortex \textit{confinement}: 
the energy of vortex-antivortex interaction \textit{grows}
(logarithmically) with the separation between them, as long as the vortex-antivortex 
spacing does not exceed London penetration depth 
$\lambda  = {\hbar}^2/[(2e)^2\mu_0E_\mathrm{J}]$.  
This condition reflects the restriction of the notion of 2D superconductivity: 
in the system with the lateral dimensions
exceeding $\lambda$, unbound vortices would present at \textit{all} finite temperatures 
(except for $T=0$) and true superconductivity, 
i.e. the state which possesses the zero resistance below some finite transition temperature is absent.

Now going to the insulating side, where $E_\mathrm{J} < E_\mathrm{C}$, 
we note that the charges of Cooper pairs in the planar Josephson junction array also interact
according to the logarithmic law~\cite{FazioSchon1991}: 
in two dimensions the Coulomb interaction between 
the charges grows logarithmically with the distance, $r$, separating them,
${\cal E}_{\mathrm{Coulomb}}\propto\ln(r/r_0)$. 
Thermal fluctuations generate excessive Cooper pairs with the charge $-2e$ each and the equal number
of ``anti-Cooper-pairs" (i.e. local deficit of Cooper pairs) 
with charges  $+2e$, to conserve the electro-neutrality.
As long as these charge separations, $r$, does not exceed 
the  electrostatic screening length, $\Lambda  = a(C/C_0)^{1/2}$,
where $C_0$ is the self-capacitance of a superconducting island (capacitance to the ground) 
and $a$ being the characteristic size of a single junction, 
the interaction energy of charges is proportional to $\ln(r/r_0)$, where in case of JJA $r_0=a$. 
Therefore, planar JJAs possess a remarkable duality between the logarithmically 
confined magnetic vortices at the superconducting 
side and the logarithmically confined Cooper pairs in the insulating state.
By the same `logarithmic token' as above the Cooper-pairs -- anti-Cooper-pairs 2D 
plasma undergoes the charge-BKT (CBKT) transition binding, 
at $T<T_{\scriptscriptstyle{\mathrm{CBKT}}}$, Cooper pairs and ``anti-pairs" 
into the neutral Cooper-pairs dipoles 
(CPD) that do not carry any charge.
The charge confinement at $E_\mathrm{J} < E_\mathrm{C}$
can be expressed in terms of the \textit{macroscopic} Coulomb blockade 
that impedes the free motion of 
the Cooper pairs and breaks loose the global phase coherence of the condensate, 
according to the uncertainty principle,
and, thus, gives rise to the superinsulating state of the JJA~\cite{VinNature}.  
In particular, at $T=0$ the magnetic vortex -- Cooper pairs charge duality 
in JJA leads to a quantum phase 
transition between superconductivity and superinsulation at the 
self-dual point where $E_\mathrm{J} \approx E_\mathrm{C}$ 
as was first noticed in~\cite{Diamantini}.  

The authors of Ref.~\cite{Diamantini} considered a two-dimensional 
Josephson junction array and showed
that the zero-temperature behavior of planar JJAs in the self-dual approximation 
is governed by an Abelian gauge theory with the periodic mixed Chern - Simmons 
term describing the charge-vortex coupling.
The periodicity requires the existence of (Euclidean)
topological excitations, which determine the quantum phase structure of the model.
Symmetry between the logarithmically 
interacting vortices and logarithmically interacting charges 
give rise to a quantum phase transition at the self-dual point.   
This is the transition between the superconducting phase and the
phase with the confined charges, dual to superconductivity  and having zero linear conductivity 
in the thermodynamic limit.  
This phase was, by analogy with the superconductors, where linear resistivity is zero, was
termed ``superinsulator"~\cite{Diamantini}.  
Therefore the quantum phase transition extensively discussed in the literature
and referred to as SIT is in fact \textit{superconductor-superinsulator transition}.

\begin{figure}[t]
\includegraphics[width=1.0\linewidth]{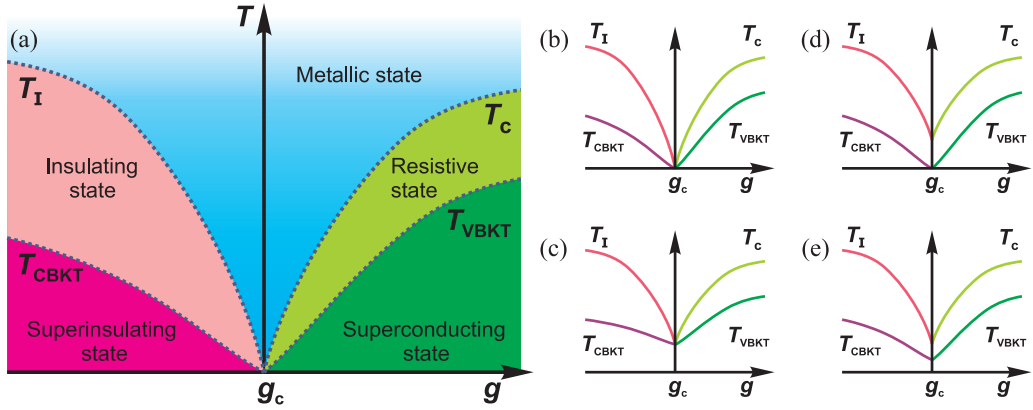}
\caption{(a) The sketch of the phase diagram for the superinsulator-superconductor transition 
in two dimensions in the close proximity to the critical point.  
 A generic diagram is plotted in coordinates temperature ($T$) - critical parameter ($g$).  
In disordered films $g$ is a conductance in the metallic phase, 
whereas in the Josephson junction arrays (JJA) it represents the ratio of coupling- and
charging energies, i.e.
$g = E_\mathrm{J}/E_\mathrm{C}$, 
where $E_\mathrm{J}$ is 
the Josephson coupling energy of the two
adjacent superconducting islands, and the charging energy $E_\mathrm{C}$, the
energy cost to transfer a Cooper pair charge 2$e$ between the neighbouring islands.
In the JJA the condition $g = g_\mathrm{c}$ means 
$E_\mathrm{J} \approx  E_\mathrm{C}$.   
At $g > g_\mathrm{c}$ the finite value of the modulus of the order parameter
 $|\Psi|$ appears at the superconducting transition temperature 
$T{_\mathrm{c}}$.  The temperature $T_{\mathrm{\scriptscriptstyle{VBKT}}}$ 
is the Berezinskii-Kosterlitz-Thouless (BKT) 
transition temperature in the 2D vortex-antivortex plasma. 
At $T{_\mathrm{\scriptscriptstyle{VBKT}}} < T < T_{\mathrm{c}}$ 
the system is in a resistive state since in this temperature range 
vortex-antivortex pairs are unbound 
and can diffuse freely breaking down the global phase coherence.  
At $T < T{_\mathrm{\scriptscriptstyle{VBKT}}}$ vortex-antivortex pairs are bound, the global phase 
coherence establishes and the superconducting state forms. 
At $g < g{_\mathrm{c}}$ and $T < T{_\mathrm{I}}$ the system turns an insulator and in
the range $T{_\mathrm{\scriptscriptstyle{CBKT}}} < T < T_{\mathrm{I}}$ it exhibits the exponentially 
low conductivity due to tunneling transfer of electric charges, 
$T_\mathrm{\scriptscriptstyle{CBKT}}$ being the temperature of the charge-BKT transition.  
At $T < T_\mathrm{\scriptscriptstyle{CBKT}}$, where the negative and positive charges 
are bound into dipoles, the system becomes superinsulating.  
The blue sector of $T>T{_\mathrm{c}}$ and $T>T_{\mathrm{I}}$ corresponds to a metallic state.
Both low-temperature states, superinsulating and superconducting, are non-dissipative.
(b)-(e) Schematic possible realizations of this phase diagram: (b) The same as (a), for the case where 
at the critical point all the transition temperatures turn zero; 
(c) All the characteristic temperatures coincide at $g=g_c$, 
but remain finite.  This situation corresponds to the \textit{first order} transition between the superconductor and superinsulator; 
(d) $T_\mathrm{\scriptscriptstyle{CBKT}}(g_{\mathrm c})=T_\mathrm{\scriptscriptstyle{VBKT}}(g_{\mathrm c})=0$, but
$T_{\mathrm{\scriptscriptstyle I}}(g_{\mathrm c})=T_{\mathrm c}(g_{\mathrm c})\neq 0$; 
and (e) $T_\mathrm{\scriptscriptstyle{CBKT}}(g_{\mathrm c})=T_\mathrm{\scriptscriptstyle{VBKT}}(g_{\mathrm c})\neq 0$ and
$T_{\mathrm{\scriptscriptstyle I}}(g_{\mathrm c})=T_{\mathrm c}(g_{\mathrm c})\neq 0$.
}
\label{fig:PhaseDiagram}
\end{figure}
Turning now to the description of the phase diagram of the JJA in the vicinity 
of the super\-con\-ductor-superinsulator transition
(Fig.~\ref{fig:PhaseDiagram}), we note first that in the 2D superconductors the symmetry 
of the phase-charge uncertainty
relation can be recast into a picture of the vortex-Cooper pair duality
(the concept of quantum superconductor-insulator transition 
as a transition between the state with pinned vortices 
and the state with pinned Cooper pairs was first introduced in the pioneering works 
~\cite{Fisher1990,Fisher1990B} as
a SIT in disordered superconducting film). 
In this context, superconductivity is viewed as an ensemble
of Cooper-pair condensate and pinned vortices.  
Inversely, the insulator consists of the Bose condensed vortices
and of the pinned Cooper pairs.
The distinctive feature of the superconductor-insulator transition 
is that this is the transition at which by its very definition one expects 
a singularity in transport properties.
Thus the symmetry in the ground states is paralleled by the symmetry 
of the electronic transport~\cite{FazioNatNV,Demler2010}
which can be mediated either by the tunnelling of the fluxons
or by the tunnelling of the Cooper pairs.  
Now the both sides of the phase
diagram of JJA near the superconductor-superinsulator transition 
(see Fig.~\ref{fig:PhaseDiagram}) can be described from the unified viewpoint:
If the linear size of the JJA is less then both characteristic screening lengths, 
the system would undergo both vortex- and charge-BKT transitions (VBKT and
CBKT) at temperatures $T_\mathrm{\scriptscriptstyle{VBKT}}\approx  E_\mathrm{J}$ 
and $T_\mathrm{\scriptscriptstyle{CBKT}}\approx  E_\mathrm{C}$, respectively
(the effects of the finite screening lengths on VBKT and CBKT transitions 
were discussed in~\cite{2DJJA_Review} and~\cite{ScrLengthCBKT}).
The VBKT transition marks the transition between the resistive 
(at $T>T_\mathrm{\scriptscriptstyle{VBKT}}$) 
and global phase coherent superconducting (at $T<T_\mathrm{\scriptscriptstyle{VBKT}}$) states.  
Accordingly, transition at $T=T_\mathrm{\scriptscriptstyle{CBKT}}$ at the insulating side 
separates the `high-temperature state,' insulator, where Cooper-pair dipoles (CPD) get unbound, 
and the `low-temperature phase,' superinsulator, where dipole excitations are bound.
 
The insulating phase at $T_\mathrm{\scriptscriptstyle{CBKT}}<T<T_{\scriptscriptstyle{\mathrm{I}}}$ 
exhibits thermally activated conductivity 
$\sigma\propto\exp(-\Delta_{\mathrm{C}}/k_{\mathrm{\scriptscriptstyle B}}T)$. 
The charge transfer is mediated by the tunnelling of Cooper pairs,
so this state is referred to as a Cooper pair insulator. 
Not too far above $T_\mathrm{\scriptscriptstyle{CBKT}}$, 
where the density of the unbound CPD is still low,
$\Delta _\mathrm{C} = E_\mathrm{C} \ln(L/a)$ (where 
 $L$ is the linear size of the array, and $L<\Lambda$)~\cite{FVB}.  
That the activation energy $\Delta _\mathrm{C}$ characterizing the insulating behavior
exhibits such an unusual and peculiar  dependence on the size of the system
reflects the fact that the test excessive charge placed locally at any superconducting island
polarizes the whole JJA~\cite{FazioSchon1991}.   
In other words, the characteristic Coulomb energy cost
associated with the excitation of an excess Cooper pair is determined by the total system capacitance  
$C_\mathrm{tot} \approx C/\ln(L/a)$~\cite{VinNature}. 
Thus, since it is the whole system that participates in building the electrostatic barrier 
impeding the free charge propagation, we refer to this peculiar Coulomb blockade effect 
as to a \textit{macroscopic} Coulomb blockade.
On the experimental side,  the size-scaling of the activation energy of the insulator 
has not been demonstrated on JJAs yet and is waiting for being revealed. 
At the same time, the scaling of the activation energy as the logarithm of 
the sample size was recently observed in disordered
InO films~\cite{Zvi2008} and in TiN films~\cite{ChargeBKT}.
We will discuss these observations in more details in the Section 1.4 below.

Upon further temperature growth, screening of an excess charge by the free unbound 
charges becomes more efficient and the characteristic activation energy reduces 
to the charging energy of a single island, $E_\mathrm{C}$.
The sketch of the phase diagram showing the possible states
in the planar JJA in $g$-$T$, coordinates, where $g=E_{\mathrm{J}}/E_{\mathrm{C}}$ 
is a critical parameter, is presented in Fig.\,\ref{fig:PhaseDiagram} and generalizes
the phase diagram proposed first by Fazio and Sch\"on~\cite{FazioSchon1991}, 
see also~\cite{Mooij1990,FazioZant2001}.
We expect that the similar diagram describes the behavior of disordered superconducting films.
In this case the role of the critical parameter is taken by the dimensionless conductance $g$, 
and the superinsulator-superconductor transition occurs at the critical conductance $g = g_c$.
The metal that forms near $g_c$ in the $T\rightarrow 0$ limit has the conductance
close to the quantum value $e^2/(2\pi\hbar)=(25.8\,\mathrm{k\Omega})^{-1}$.
The corresponding distinct state has been detected
via applying strong magnetic field completely destroying superconductivity
in disordered films of Be, TiN, and InO, and is often referred to as
``quantum metallicity"~\cite{TBJETPL,TiNQM,ButkoAdams,SITInOKapitulnik,TiNHA}.

An  important comment is in order.  
Two-dimensional systems display perfect superconducti\-vi\-ty-superinsulation duality
because of the logarithmic interaction of charges in 2D.  
One can ask whether this duality consideration 
can be extended onto three dimensional systems and whether one 
can expect superinsulating behavior in 3D.
One can see straightforwardly that while 3D Abrikosov vortices 
still interact according to the logarithmic law,
the charges follow the conventional 3D Coulomb law and therefore charge-anticharge 
confinement is absent in 3D.
So absent is the 3D superinsulating state.    

\section{Two-dimensional universe in superconducting films}

As we have already discussed, Josephson junction arrays with the junction capacitances 
well exceeding their respective capacitances to the ground are two-dimensional 
with respect to their electrostatic behavior
since most of the electric force lines remain trapped within the junctions themselves. 
When considering behavior of homogeneously disordered superconducting films, 
the question can arise, 
to what extent they can be described by physics derived for JJAs. 

The analogy between the planar JJA and the critically disordered thin superconducting 
films can be perceived 
from the fact that the experimentally observed in TiN films magnetic-field dependences
of the activation energy, $\Delta_{\mathrm C}(B)$, and the threshold voltage 
$V_{\mathrm{\scriptscriptstyle{T}}}(B)$~\cite{TiNSIT}
are remarkably well described 
by formulas obtained in the framework of JJA model~\cite{FVB}.  
This hints that in the critical vicinity of the SIT a film can be viewed 
as an array of superconducting islands coupled by the weak links.
This incipient conjecture of the early work~\cite{Zvi1994} discussed further in
Refs.~\cite{GhosalPRL,Valles1999,GhosalPRB,DubiNat,Trivedi2011,TiNSIT,TiNPhysC,Zvi2008,TiNHA,
ShaharSteps,KalokLT26,VinNature} 
that near the disorder-driven SIT the electronic phase separation occurs leading to
formation of the droplet-like (or island-like) texture, superconducting 
islands coupled via the weak links, 
was supported by recent scanning tunnelling spectroscopy (STS) 
findings~\cite{BaturinaSTM08,InOSTM_NP}.
In general, formation of regular textures and, in particular, the spontaneous
self-organization of electronic nanometer-scale structures, due to the existence 
of competing states is ubiquitous
in nature and is found in a wealth of complex systems and physical 
phenomena ranging from 
 magnetism~\cite{Khomskii1984,Khomskii1999,Khomskii2001,Khomskii2008,Khomskii2009},
superconductivity and superfluidity to liquid crystals, see Ref.~\cite{Dagotto} for a review.
The arguments that the long-range fields -- for example, elastic or Coulomb --
can promote phase separation were given in Ref.~\cite{Islands} where
the island texture due to elastic interaction between the film and the substrate was found.
More refined calculations for the JJA model~\cite{Syzranov2010} showed 
that in the presence of the long-range Coulomb forces,
the SIT could become a first order transition supporting 
the idea of possible electronic phase separation at the SIT.

\begin{figure}[t]
\includegraphics[width=0.6\linewidth]{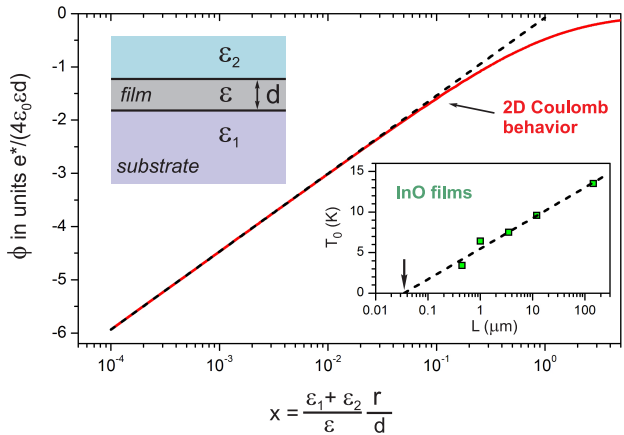}
\caption{  The electrostatic potential (shown as the solid line) induced by the charge $-e^{\star}$ 
in the film with the dielectric constant $\varepsilon$
placed in between of two media with the dielectric constants 
$\varepsilon_1$ and $\varepsilon_2$, $\varepsilon_1+\varepsilon_2\ll \varepsilon$
(see the left inset) as function of the logarithm of 
the reduced charge separation $x=[(\varepsilon_1+\varepsilon_2)/\varepsilon](r/d)$ 
according to Eq.\,(\ref{2DCoulomb})
in the units of $e^{\star}/(4\varepsilon_0\varepsilon d)$.  
The dashed line shows the logarithmic asymptote of the potential offering 
a perfect approximation in the region $d\ll r\ll \varepsilon d/(\varepsilon_1+\varepsilon_2)$.  
The right inset displays the fit of the 
data on the dependence of the activation energy of the InO films on the size  of the sample 
from Ref. ~\cite{Zvi2008} by
Eq.\,(\ref{2Dlog}) (the dashed line) according to~\cite{FVB}. 
The intersection of the line with the $\mathrm{x}$-axis gives 
the low-distance cutoff  of 35\,nm, which agrees
fairly well with the InO film thickness 25\,nm quoted in~\cite{Zvi2008}.  
The slope of the fit allows then for determination of $\varepsilon$.
}
\label{fig:2DCoulomb}
\end{figure}

Furthermore, while at the superconducting side of the transition vortices 
interact logarithmically, thus guaranteeing all the BKT physics to occur, 
provided the size of the system remains less than the magnetic screening length, 
the question arises whether the charges in the same film but at the insulating side would also
interact according to the logarithmic law.  
In simple words, in order for the film to demonstrate the 
2D behavior, the electric force lines are to be trapped within the film over 
the appreciable distance.
One can derive from the fundamentals of electrostatics that if the test charge is placed 
within the dielectric film of the thickness $d$ and with the dielectric constant 
$\varepsilon$, the characteristic length over which the electric field remains trapped 
within the film is of about $\varepsilon d$.  
The quantitative description of the Coulomb interaction in a thin film 
was first given by Rytova~\cite{Rytova1967}, the logarithmic asymptotic
of the solution and its implications were discussed by Chaplik and Entin~\cite{ChaplikEntin1971},
and a refined calculation for a film 
with the large dielectric constant $\varepsilon$
sandwiched between the two dielectric media with the dielectric 
constants $\varepsilon_1$ and $\varepsilon_2$
(see the left inset in Fig.\,\ref{fig:2DCoulomb}) 
was done by Keldysh~\cite{Keldysh1979}.  
The dependence of the electrostatic potential of the charge $-e^{\star}$ 
on the distance $r\gg d$,  is given by (see Fig.~\ref{fig:2DCoulomb}):
	\begin{equation}\label{2DCoulomb}
		\phi(\vec{r})=-\frac{e^{\star}}{4\varepsilon_0\varepsilon d}
		\left[{\cal H}_0\left(\frac{\varepsilon_1+\varepsilon_2}{\varepsilon}\frac{r}{d}\right)
		-{\cal N}_0\left(\frac{\varepsilon_1+\varepsilon_2}{\varepsilon}\frac{r}{d}\right)\right]\,,
	\end{equation}
where ${\cal H}_0$ and ${\cal N}_0$ are the Struve and Neumann 
(or the Bessel function of the second kind) functions, respectively.
This formula corresponds to the choice $\phi(r)\to 0$ as $r\to\infty$. 
In the region $d\ll r\ll \Lambda$, where 
\begin{equation}\label{eq:Lambda}
		\Lambda=(\varepsilon d)/(\varepsilon_1+\varepsilon_2)\,
	\end{equation}
is the electrostatic screening length, the electric force lines are trapped by the film, and 
the interaction energy between the
charges $e^{\star}$ and $-e^{\star}$ obtained from Eq.\,(\ref{2DCoulomb})
acquires the form
	\begin{equation}\label{2Dlog}
		V(\vec{r})= \frac{(e^{\star})^2}{2\pi\varepsilon_0\varepsilon d}\ln\left(\frac{r}{d}\right)\,,
		\,\,\,\, d\ll r \ll \Lambda\,.
	\end{equation}

The universal behavior of the electrostatic potential as a function 
of the reduced coordinate $x=[(\varepsilon_1+\varepsilon_2)/\varepsilon](r/d)\equiv r/\Lambda$
is shown in the Fig.\ref{fig:2DCoulomb} by the solid line. 
The dashed line shows the logarithmic approximation valid in the spatial region
$d\ll r\ll \Lambda$.   
The right inset displays the fit of the 
experimental data on the dependence of the activation energy of the InO films 
on the size of the sample from~\cite{Zvi2008} by
Eq.\,(\ref{2Dlog}) (the dashed line) according to~\cite{FVB}. 
The intersection of the line with the $\mathrm{x}$-axis gives 
the low-distance cutoff of 35\,nm which agrees
fairly well with the InO film thickness 20$\div$25\,nm quoted in~\cite{Zvi2008}.  
The slope of the fit allows then for determination 
of $\varepsilon$ and, taking where $e^{\star}=2e$, yields $\varepsilon=2.3\cdot 10^3$.  
One can then estimate the charging energy of an `effective single junction'
$E_{\mathrm C}=(2e)^2/(4\pi\varepsilon_0\varepsilon d)$ and find the value for the 
$T_{\scriptscriptstyle{\mathrm{CBKT}}}\simeq E_{\mathrm C}/k_{\scriptscriptstyle{\mathrm{B}}}\approx 0.8$\,K for InO films.
Using the data for the size-dependence of the activation 
energy from~\cite{ChargeBKT} we can carry out the similar estimates  
for TiN films as well.  
The intersection of the linear fit similar to that shown in the right 
inset of Fig.\ref{fig:2DCoulomb}, one finds
$d=3$\,nm which is fairly close to their nominal thickness of 5\,nm.  
Then the slope of the fit gives $\varepsilon\simeq 4\cdot 10^5$, and accordingly,
$T_{\scriptscriptstyle{\mathrm{CBKT}}}\simeq E_{\mathrm C}/k_{\scriptscriptstyle{\mathrm{B}}}\approx 0.06$\,K.  
The latter estimate is in a pretty  good agreement with the experimental 
findings by~\cite{ChargeBKT,KalokLT26} that gave 
$T_{\scriptscriptstyle{\mathrm{CBKT}}}^{\scriptscriptstyle{\mathrm{IV}}}\lesssim 60$\,mK 
for the TiN film.
One can further estimate the electrostatic screening length $\Lambda$.
Using the parameters determined from the fit and 
$\varepsilon_1=4$ for SiO$_2$ and $\varepsilon _2=1$, 
one finds  $\Lambda\simeq 240$\,$\mu$m. 
This macroscopic value of $\Lambda$ implies that
all the TiN samples used in experiments~\cite{VinNature,TiNHA,TiNSIT,TiNPhysC} having
50\,$\mu$m width fall well into a domain of validity of a two-dimensional electrostatics. 
One of the conclusions that follows from our estimates is 
that the condition $\varepsilon_1+\varepsilon_2\ll \varepsilon$
is an important requirement for the observation of the superinsulating behavior; 
so one has to take a special care
of taking the substrate with a reasonably low $\varepsilon_1$.   
Another comment is 
that while our results allow for the fairly reliable determination of $\varepsilon$ 
and are remarkably self-consistent,
yet the direct measurements of the film dielectric constant in the critical vicinity of the 
SIT are highly desirable.

An important feature of the evaluated parameters is that at the first sight the dielectric constant,
especially if talking about the `conventional' bulk materials, seem to come out pretty high.  
However, it ceases to be surprising once we recollect that the experiments 
were carried out in the close proximity to the SIT.
Indeed, in  a seminal paper of 1976, Dubrov \textit{et al}~\cite{Dubrov}
presented two complimentary considerations based on the percolation theory 
and on the theory of the effective media
showing that the static dielectric constant should become infinite 
at the metal-insulator transition (see also~\cite{Stauffer}).
The authors were motivated by the experimental finding 
by Castner and collaborators~\cite{Castner1975} who 
revealed the onset of a divergence in the static dielectric constant of 
\textit{n}-type silicon at the insulator-metal transition 
(this phenomenon known as \textit{polarization catastrophe} 
was predicted as early as in 1927~\cite{Herzfeld1927}).
Furthermore, Dubrov and collaborators investigated numerically and experimentally the model system, 
the square conducting network, where each bond was a parallel resistor-capacitor circuit.
Breaking randomly the connections between the elements they demonstrated 
that near the percolation transition 
that the dielectric constant diverges near the percolation transition in a power-law fashion.
\begin{figure}[t]
\includegraphics[width=1.0\linewidth]{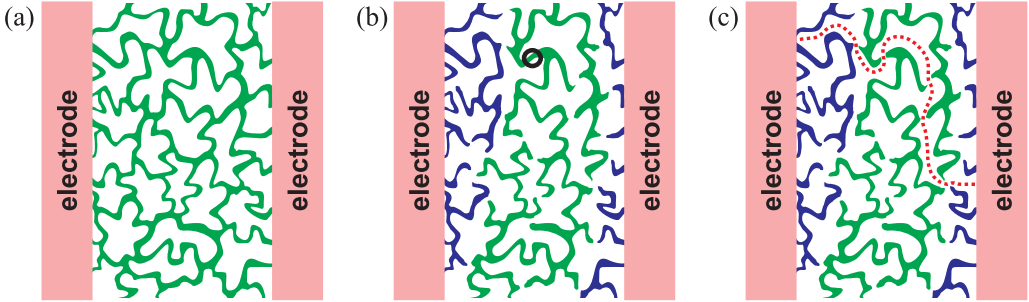}
\caption{ 
A sketch of the electronic phase separation in a disordered film.  
(a) An insulating state, where superconducting regions 
(shown as white areas) are separated 
by the insulating spacers. 
(b) An insulating state on the very verge of the percolation transition:
the green stripe highlights the ``last" or ``critical" 
insulating interface dividing two large superconducting clusters.  
The circle marks the critical (``last") insulating ``bond." 
(c) The same system at the extreme proximity to the SIT 
on the superconducting side.  
As soon as the ``critical" insulating bond is broken, 
the path connecting electrodes through the superconducting 
clusters shown by the dotted line appears and 
the film turns superconducting.  
The system capacitance $C$ on the insulating side is proportional to the length of the insulating 
interface separating two adjacent superconducting clusters, which grows near the transition 
as $b^{1+\psi}$, $\psi > 0$, i.e. faster than $b$.  
Thus the dielectric constant  $\varepsilon\sim C/b$ diverges upon approach to 
superconductor-superinsulator transition.
}
\label{fig:filmSIT}
\end{figure}
In what follows we present the arguments, following the the considerations of Ref.~\cite{Dubrov}
 by which the two-phase systems that undergo 
the transition between the conducting and non-conducting states should
exhibit the diverging dielectric constant when approaching this transition.  
A qualitative picture of this phenomenon is illustrated in Fig.\,\ref{fig:filmSIT}.
The Panel (a) shows the spatial electronic structure of an
insulating film in a close proximity to the percolation-like 
superconductor-insulator transition~\cite{Shimshoni1998}
approaching it from the insulating side.
The white areas denote the superconducting regions which are separated
from each other by the dark insulating areas.
The Panel (b) displays the same film but ``one step before"
the transition: upon breaking the ``last" insulating separation 
(the circled segment in the panel)
between the superconducting clusters, the path connecting electrodes
through superconducting clusters emerges and the film becomes superconducting as shown in Panel (c).
The dielectric constant is proportional to the capacitance of the system, $C$, which in its turn,
is proportional to the total length of the line separating 
the adjacent critical superconducting clusters.
If we let the linear size of the superconducting cluster be $b$, this length would grow as $b^{1+\psi}$ 
with the exponent $\psi > 0$ (see Ref.~\cite{Stauffer}).  Then the dielectric constant 
$\varepsilon\propto C/b\propto b^{\psi}$.  As the size of the critical cluster diverges on approach the
transition, so does the dielectric constant $\varepsilon$
(for more refined derivation of this divergence see~\cite{Dubrov}).
Such a behavior, which is called the ``dielectric catastrophe" and 
is very well known in semiconductor physics,
 has been observed experimentally (see, for example, Ref.~\cite{Rosenbaum},
where the $\varepsilon$ divergence was observed near the metal-insulator transition).  
Therefore one would expect that in the critical vicinity
of the superconductor-insulator transition, the dielectric constant grows large enough so that 
the Coulomb interaction between the charges became logarithmic over the appreciable scale comparable
to the size of the system, 
and the 2D plasma of CPD excitations experiences  the charge-BKT transition 
leading to formation of the low-temperature superinsulating state.

We now turn to the structure of the phase diagram 
of Fig.\,\ref{fig:PhaseDiagram}.
At the very transition the dielectric constant $\varepsilon\to\infty$, and, accordingly,
the charging energy $E_{\mathrm{C}}\to 0$.
Therefore, at the self-dual point, $g=g_{\mathrm{c}}$, both characteristic temperatures vanish,
$T_{\mathrm{\scriptscriptstyle{CBKT}}}\equiv T_{\mathrm{\scriptscriptstyle{VBKT}}}=0$.
Upon decreasing $g$ and departing from the transition to the insulating side, 
$\varepsilon$ decreases and
 $T_{\mathrm{\scriptscriptstyle{CBKT}}}\simeq E_{\mathrm{C}}$ 
grows while going deeper into the insulating side as shown in Fig.\,\ref{fig:PhaseDiagram}.  
Note that the activation energy $\Delta_{\mathrm{C}}=E_{\mathrm{C}}\ln(L/d)$
also grows upon decreasing $g$ (or increasing the resistance of the film) 
as $(g_{\mathrm{c}}-g)^{\nu}$, with the exponent $\nu>0$ and so does
the crossover line $T_{\mathrm{I}}\simeq\Delta_{\mathrm{C}}$.  
The width of the insulating domain in the phase diagram confined 
between $T_{\mathrm{\scriptscriptstyle{CBKT}}}$ and $T_{\mathrm{I}}$ increases as well.
Upon further decrease in $g$ and the corresponding decrease 
in $\varepsilon$, the electrostatic screening length $\Lambda\propto \varepsilon d$ shrinks, 
and as it drops down appreciably below the sample size, the superinsulating domain
at the phase diagram ceases to exist.

Having discussed the behavior of the films at the insulating side, we now review, for completeness the
superconducting side.
Suppression of the superconducting transition temperature, $T_{\mathrm{c}}$, by disorder in thin
superconducting films can be traced back to pioneering 
works by Shal'nikov \cite{Shalnikov38,Shalnikov40}, 
where it was noticed for the first time, that  $T_c$ decreases with the decrease of the film thickness.
The next important step was made by Strongin and collaborators~\cite{Strongin70}, who had found that
the $T_{\mathrm{c}}$ correlates with the sheet resistance  
$R_{\square}=3\pi^2\hbar/(e^2k_{\scriptscriptstyle{\mathrm{F}}}^2\ell d)=1/g$,
much better than with the thickness of the film or its resistivity, 
and thus has to serve as a measure of disorder in the film.

\begin{figure}[t]
\includegraphics[width=0.8\linewidth]{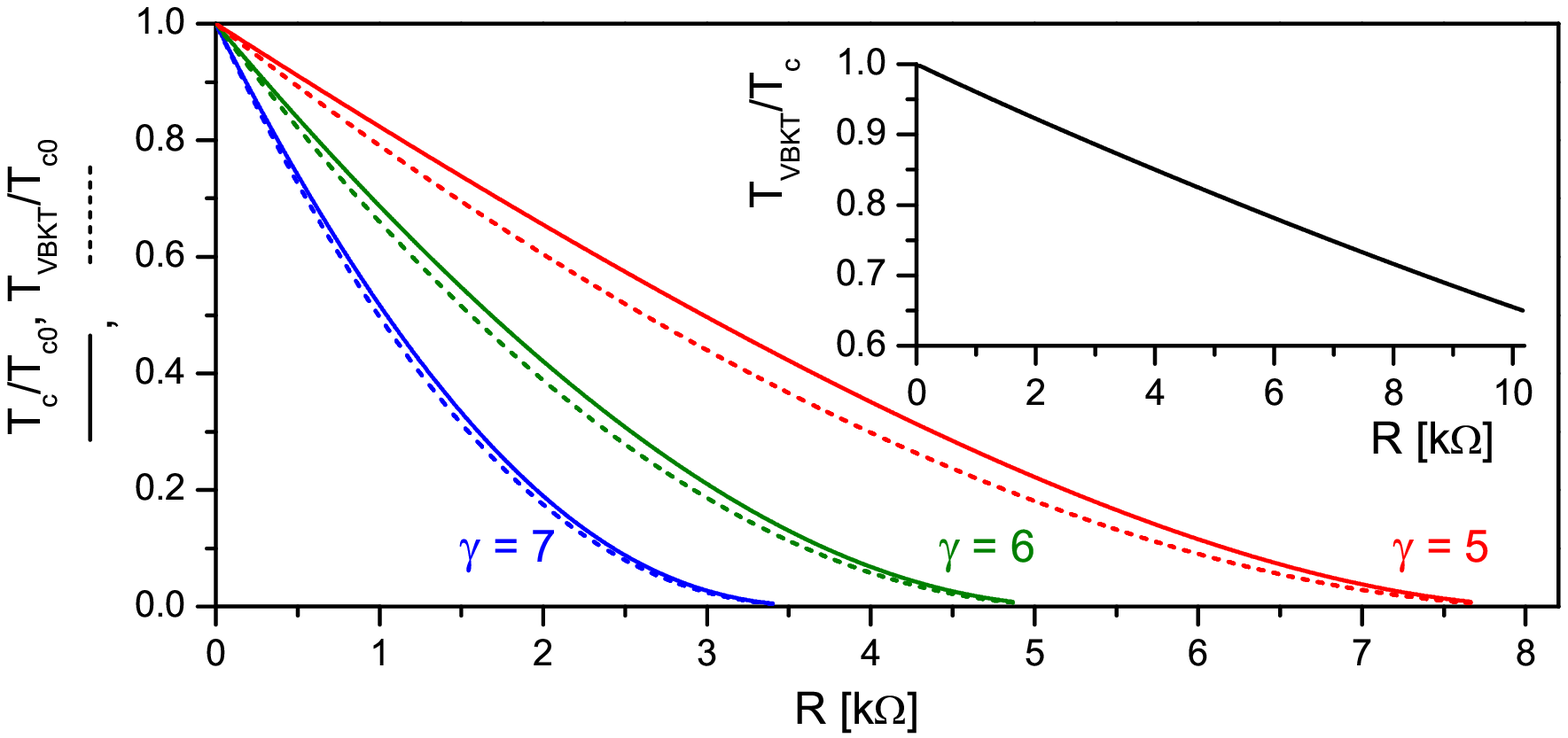}
\caption{Reduced critical temperature $T_{\mathrm{c}}/T_{\mathrm{c0}}$ 
vs. sheet resistance according to Eq.\,(\ref{FinTc}) (solid lines) for different parameters $\gamma$.
The values $\gamma= 5,\,6,\,7$ correspond to 
$\tau=10.3\cdot 10^{-15}$\,sec, $4.8\cdot 10^{-15}$\,sec, and $1.4\cdot 10^{-15}$\,sec,
respectfully, at $T_{\mathrm{c0}}=5$\,K.
The inset shows $T_{\scriptscriptstyle{\mathrm{VBKT}}}/T_{\mathrm{c}}$ vs. resistance BMO relation
calculated from Eqs.\,(\ref{BMOfunc1}) and (\ref{BMOfunc2}), which is used to find 
$T_{\scriptscriptstyle{\mathrm{VBKT}}}/T_{\mathrm{c0}}$ shown by the dashed lines.
}
\label{fig:TcTVBKT}
\end{figure}

Since then, numerous experimental works revealed a drastic suppression
of $T_c$ in various superconducting films with growing the sheet resistance, such as 
Pb~\cite{Strongin70,Dynes1986,HavGold,LiuGoldman,Xiong1995}, 
Al~\cite{LiuGoldman}, 
Bi~\cite{Strongin70,HavGold,LiuGoldman,Marcovic98,Valles1999}, 
W-Re alloys~\cite{Raffy83}, MoGe~\cite{Graybeal84,Rogachev2012}, 
InO~\cite{Hebard1985,ShaharOvadyahu,VFGIns}, Be~\cite{Be1}, 
MoSi~\cite{Okuma1998},
Ta~\cite{Yoon2006}, NbSi~\cite{Aubin2008},
TiN~\cite{TiNJapan2000,TBJETPL,Hadacek,TiNSIT,TiNPhysC,BaturinaSTM08,Baturina2011},
  and PtSi~\cite{OtoPtSi1994}, to name a few. 
  The observed behavior appeared to be in a good quantitative accord with the
theoretical predictions by Maekawa and Fukuyama~\cite{Maekawa}, 
who first connected quantitatively the reduction in $T_{\mathrm c}$
as compared to its bulk value $T_{\mathrm{c0}}$ with $R_{\square}$,
and with those of the subsequent work by Finkel'stein~\cite{Finkelstein}, 
who came up with the comprehensive formula for $T_{\mathrm c}(R_{\square})$:
\begin{equation}
\ln \left(\frac{T_{\mathrm{c}}}{T_{\mathrm{c0}}}\right)=
\gamma+\frac{1}{\sqrt{2r}}
\ln\left(\frac{1/ \gamma + r/4 - \sqrt{r/2}}{1/ \gamma + r/4 + \sqrt{r/2}}\right)\,,
\label{FinTc}
\end{equation}
where $r=R_\Box e^2/(2\pi ^2 \hbar)$ and 
$\gamma=\ln[\hbar/(k_{B}T_\mathrm{{c0}}\tau)]$.  
The reduction of $T_{\mathrm{c}}$ with the increasing sheet resistance 
according to this formula is shown 
in Fig.\,\ref{fig:TcTVBKT} for several values of the product $T_{\mathrm{c0}}\tau$.

The physical picture behind suppressing superconductivity by disorder
is that in quasi-two-dimensional systems disorder inhibits
electron mobility and thus impairs dynamic screening of the Coulomb interaction.
This implies that disorder enhances effects of the Coulomb repulsion between electrons which, 
if strong enough, breaks down Cooper pairing and destroys superconductivity.
Importantly, according to~\cite{Finkelstein},
the degree of disorder at which the Coulomb repulsion would balance
the Cooper pair coupling is not sufficient to localize normal carriers.  
Therefore, at the suppression point superconductors transforms into a metal.
The latter can transform into an insulator upon further increase of disorder.
Therefore this mechanism, which is referred to as a fermionic mechanism,
results in a sequential superconductor-metal-insulator transition~\cite{Finkelstein}.
However, several experiments~\cite{BaturinaSTM08,Hadacek} 
demonstrated that in the vicinity of the SIT
the dependence $T_{\mathrm{c}}(R_{\square})$ follows perfectly well the Finkel'stein's 
formula.  We would like to reiterate here that in spite of the fact that 
Finkelstein's theory treats superconductor-metal transition, and moreover,
the films in the critical region may develop strong mesoscopic fluctuations~\cite{Feig2005},
the experimentally observed $T_{\mathrm{c}}$ near the SIT where $T_{\mathrm{c}}\to 0$,
is still perfectly fitted by formula~(\ref{FinTc}).

At $g=g_{\mathrm{c}}$ the superfluid density of the Cooper 
pair condensate is zero so at the transition point $T_{\mathrm{\scriptscriptstyle{VBKT}}}=0$.
An increase in $g$ suppresses effects of quantum fluctuations near the transition 
and, correspondingly,
$T_{\mathrm{\scriptscriptstyle{VBKT}}}$ increases.  
Using the dirty-limit formula which relates the two-dimensional
magnetic screening length to the normal-state resistance $R_{\mathrm N}$,
Beasley, Mooij, and Orlando (BMO)~\cite{Beasley1979} 
proposed the universal expression for ratio $T_{\scriptscriptstyle{\mathrm{VBKT}}}/T_{\mathrm c}$
(see inset to Fig.\,\ref{fig:TcTVBKT}): 
\begin{equation}
\frac{T_{\scriptscriptstyle{\mathrm{VBKT}}}}
{T_{\mathrm c}}f^{-1}\left(\frac{T_{\scriptscriptstyle{\mathrm{VBKT}}}}{T_{\mathrm c}}\right)
=0.561\frac{\pi^3}{8}\left(\frac{\hbar}{e^2}\right)
\frac{1}{R_{\mathrm N}}\,,
\label{BMOfunc1}
\end{equation}
\begin{equation}
f\left(\frac{T}{T_{\mathrm c}}\right)
=\frac{\Delta(T)}{\Delta(0)}\tanh\left[\frac{\beta \Delta(T)T_{\mathrm c}}{2\Delta(0)T}\right]\,,
\label{BMOfunc2}
\end{equation}
where $\Delta(T)$ is the temperature dependence of the superconducting gap and
parameter $\beta = \Delta(0)/(k_{\scriptscriptstyle{\mathrm B}} T_{\mathrm c})$.
As it was shown in studies on InO~\cite{HebardKotliar1989} and TiN~\cite{Baturina2011} 
the BMO formula
well agrees with the experiment and correctly describes a decrease in the ratio
 $T_{\scriptscriptstyle{\mathrm{VBKT}}}/T_{\mathrm c}$
with increasing disorder, provided
the proper choice of $R_{\mathrm N}$  and $\beta$ based on the experimental data is made.
Two comments are in order.  
Note that the merging tails of $T_{\scriptscriptstyle{\mathrm{VBKT}}} $ and $T_{\mathrm{c}}$,
where they both tend to zero in the Fig.\,\ref{fig:TcTVBKT}) is, in fact, the image of the same transition
point of the phase diagram of Fig.\,\ref{fig:PhaseDiagram} where the curves were plotted as functions of
the conductance.  
Second, we would like to emphasize here that the behavior of the superconducting films 
near the SIT is fully controlled by the unique parameter, the sheet resistance.

To conclude here, we demonstrated that close to the superconductor-superinsulator transition
the superconducting and insulating sides of the phase diagram are the mirror images of each other
with the correspondence between the vortex- and charge-BKT transition.  
As we have mentioned above, this duality is paralleled by the 
symmetry of transport properties: 
the VBKT current-voltage ($I$-$V$) characteristics, $V\propto I^{\alpha}$, is mirrored by the CBKT
$I\propto V^{\alpha}$ power-law behavior with 
the interchanging current and voltage~\cite{VinNature}.  
Accordingly, the critical current of a superconducting state 
setting the upper bound for the loss-free currents
which superconductor can still support maps onto the threshold voltage which marks 
the dielectric breakdown of the superinsulating state and appearance of the finite conductivity.  
This duality is illustrated by recent studies of the VBKT transition 
in TiN films ~\cite{Baturina2011} which revealed the classical
BKT transport behavior: the power-law $V\propto I^{\alpha}$ current-voltage characteristics with the 
exponent $\alpha$ jumping from $\alpha=1$ to $\alpha=3$ 
at $T=T_{\scriptscriptstyle{\mathrm{VBKT}}}$, 
in accordance with the Halperin-Nelson prediction~\cite{HalperinNelson1979}, 
and then rapidly growing with the further decrease
in temperature.  
The measurements of the current-voltage ($I$-$V$)
dependences on the same material but at the insulating side 
of the transition~\cite{ChargeBKT,KalokLT26}, 
revealed the dual $I$-$V$ characteristics which evolve upon lowering the temperature 
in precisely BKT-like manner, going from ohmic to power-law, $I\propto V^{\alpha}$, behavior with 
with the exponent $\alpha$ switching from $\alpha=1$ to $\alpha=3$ 
at $T=T_{\scriptscriptstyle{\mathrm{CBKT}}}$
and then rapidly growing from $\alpha\approx 3$ to well above the unity 
with the decreasing temperature which is again an inherent feature of the BKT transition.  

The dual similarity between the two states manifests 
itself further in that both states are the loss-free ones as the condition 
for the Joule loss $P=IV=0$ holds both in superconductors and superinsulators.
This and the mirror correspondence between the $I$-$V$ 
curves hold high potential for applications, which we will discuss below in the 
Section 7.
\section{Microscopic mechanism of conductivity in the insulating state}

The beginning of extensive studies of role of strong correlations 
and effects of disorder in the insulating behavior
can be traced back to the classical 
works~\cite{Mott1937,Mott1949,Anderson1958,Kohn1964,Abrahams1979},
see also~\cite{ImadaReview,Brandes2003}. 
Recent years have seen an explosive growth of investigations 
of the interplay between the many-body and disorder
effects leading to formation of non-conventional disorder-induced insulators
the properties of which are intimately connected with the localization 
phenomena~\cite{VinLark,Basko2006,Gornyi,BLV,VinNature,Falko2009,AAS2010}.
Yet, at present we are able to describe the tunnelling transport in the insulators only with the 
exponential accuracy  -- using euristic considerations -- as an Arrhenius 
or Mott and/or Efros-Shklovskii  behaviour.  
The challenge posed by experimental observations, 
the transition from an activated behaviour in the insulating state 
to the practically
complete suppression of the tunnelling current current in the superinsulator remains unmet.  

\begin{figure}
\includegraphics[width=0.5\linewidth]{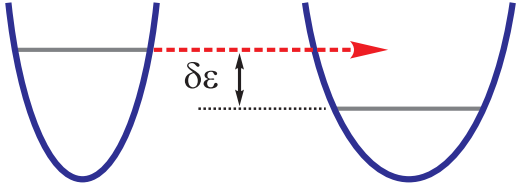}
\caption{A sketch of tunnelling between two mesoscopic potential well.  
Since the energy levels in these wells are, in general vary due to
structural disorder, quantum mechanical tunnelling is possible if 
and only if the tunnelling particle would emit (or absorb)
some bosonic excitation which would accommodate the energy difference 
$\delta\varepsilon$ between the respective energy levels.}
\label{tunneling}
\end{figure}

In what follows we will discuss one of the aspects of tunnelling transport 
in disorder-induced insulators which 
may be the key to understanding the dynamic insulator-superinsulator transition. 
Namely, in the case of disorder-induced Cooper pair insulators
the charge transfer can be viewed as a tunnelling between the localized 
Cooper pair sites possessing the essentially different energy levels, see Fig.\,\ref{tunneling}.  
The tunnelling is possible then  only in the presence of some 
energy relaxation mechanism which is able to accommodate 
the energy differences between the Cooper pair states at the 
neighbouring localize sites hosting the tunnelling process, see Fig.\,\ref{tunneling}.  

One can propose that an appropriate theory of transport in the insulating state should 
be constructed via the incorporation of the ideas of relaxation physics into
a general model of  transport in granular materials. As a starting point one can take
a model that had already been successfully used by Efetov ~\cite{Efetov1980}
to demonstrate the very existence of the disorder-driven SIT.
The customary relaxation mechanism in `conventional' conductors is the energy exchange 
between the tunnelling charge carriers and phonons, comprising a thermal bath.  
In granular materials, at low temperatures however, the role of the major relaxation processes,
ensuring the tunnelling charge transfer, is taken by emission 
(and/or absorption) of dipole excitations.
Importantly, these excitations are the same particles that mediate the charge transfer.
In the Cooper-pair insulators the dipoles are thus made up of the local excess (-2$e$) 
and local deficit (+2$e$) in the Cooper-pairs number
and form the bosonic
 environment~\cite{FVB,VinNature,BLV,Lopatin2007,CVB2009,CVBPhysC2010,CVB2011NATO}. 
The important features of this dipole excitations environment is that the dipoles 
are generated in the process of tunnelling and that 
the dipole  environment possesses an infinite number of degrees of freedom, 
and as such it efficiently takes away the 
energy from the tunnelling particles and plays the role of the thermostat itself.  
Eventually, dipole excitations relax energy to the phonon bath.
A theory of such a sequential, \textit{cascade} relaxation in 1D granular arrays 
was developed in~\cite{CVB2009,CVBPhysC2010,CVB2011NATO}.
One of the important conclusions of this theory is that the interaction of the environment 
with the infinite number of degrees of freedom  with disorder gives rise to
broadening the levels of the tunnelling carriers.
Another important result is that as soon as the spectrum of the environment 
excitations becomes gapped,
the relaxation vanishes and the tunnelling current becomes completely suppressed.
One can thus conjecture that it is the localization transition of the environmental excitation spectrum 
that marks the appearance of the gap in the local spectrum of environmental excitations 
and thus the suppression of the tunnelling charge transfer.

In a two-dimensional array the dipole excitations
form the two-dimensional Coulomb plasma, such a transition, and, respectively, the
suppression of the relaxation rate takes place at the temperature 
of the charge BKT transition, 
$T_{\mathrm{\scriptscriptstyle{CBKT}}}\simeq \bar{E}_{\mathrm{C}}$, 
where $\bar{E}_{\mathrm{C}}$ is the average charging energy of a single granule.
Below this temperature charges and anti-charges get bound into the neutral CPD, 
the glassy state forms, the gap in the \textit{local} density of states of the environmental 
excitation spectrum appears
and the tunnelling current vanishes~\cite{FVB,VinNature,CVB2009,CVBPhysC2010,CVB2011NATO}.
Although the detailed analytical calculations in two dimensions are not available at this point, 
the conjecture that the microscopic mechanisms of suppression of conductivity 
in one- and two dimensions is of the similar nature
and are due to the appearance of the gap in the local density 
in the CPD excitation spectrum (i.e. localization
in the energy space) promises to offer a route towards a quantitative microscopic mechanism behind 
the formation of the superinsulating state.
 
Recently, the transition from the insulating to superinsulating phase
at low but finite temperature due to suppression of the energy relaxation
was also found for the model of superconductor-insulator transition
on the Bethe lattice~\cite{Ioffe2010,FIM2010}.

Remarkably, the employed concept of the relaxation mediated 
by the emission/absorption of dipole excitation explains also
the simultaneous (i.e. occurring at about the same temperature)
suppression of both, Cooper pairs and the normal excitations tunnelling currents.  
However, the detailed study of this appealing topic still remains a challenging task.

\section{Conductivity of the Cooper-pair insulator and superinsulator: qualitative consideration}

Based on the ideas of the cascade relaxation discussed in the previous section we can offer 
a qualitative description of the temperature evolution of  conductivity at the insulating side. 
A crude estimate can be obtained following heuristic considerations of Ref.~\cite{VinNature}.
The tunnelling current in an array of superconducting granules can be written in the 
following form~\cite{Ingold1991,CVB2009,CVB2011NATO}  
    \begin{equation}
        I\propto\exp(-E/W)\,,
        \label{eq:SI-current}
    \end{equation}
where $E$ is the characteristic energy barrier controlling the charge transfer 
between the granules, $W=\hbar/\tau_{\scriptscriptstyle{\mathrm{W}}}$, and 
$\tau_{\scriptscriptstyle{\mathrm{W}}}$ is the relaxation time,
i.e. the time characterizing the rate of the energy exchange between
the tunnelling charges and the environment.
%
One would expect that the rate of relaxation is proportional to the 
density of the CPD excitations, which, in its turn, 
can be taken proportional to the Bose distribution function (CPD excitations are bosons).
The width of the local energy gap is $\bar{E}_{\mathrm{C}}$, therefore one 
can take 
        \begin{equation}
             \frac{\hbar}{\tau_{\scriptscriptstyle{\mathrm{W}}}}
             \simeq\frac{\bar{E}_{\mathrm C}}{\exp({E}_{\mathrm C}/T)-1}\,.
             \label{eq:W}
        \end{equation}
In other words, the relevant energy scale characterizing
the tunnelling rate is the energy gap that enters with the weight equal to
the Bose filling factor describing the probability of exciting the unbound charges.
Above the charge BKT transition Eq.\,(\ref{eq:W}) gives $W\simeq T$;
this is nothing but the equipartition theorem telling us that the number 
of the unbound charges is merely proportional to $T/{E}_{\mathrm C}$).  
To complete the estimate we have to evaluate the characteristic energy $E$.
In a 2D JJA or in a 2D disordered film in the vicinity of the SIT, 
where the electrostatic screening length $\Lambda$ is large
and exceeds the linear sample size, $L$, the characteristic energy
$E={E}_{\mathrm C}\ln(L/a)$, where $a$ is the size of a single Josephson junction.
One realizes that well above  $T_{\scriptscriptstyle{\mathrm{CBKT}}}$, the logarithmic charge interaction is
screened so that $\Lambda\approx a$ and $E$ is reduced to ${E}_{\mathrm C}$.
In this temperature region the system exhibits the `bad metal' behavior.
However, if one is not too far from the charge-BKT transition, 
$T\gtrsim T_{\scriptscriptstyle{\mathrm{CBKT}}}$, 
one still has $\Lambda >L$ and
conductivity acquires Arrhenius thermally activated form 
with the activation energy that scales as $\ln(L/a)$:
    \begin{equation}
        \sigma\propto\exp[-{E}_{\mathrm C}\ln(L/a)/T]\,, \,\,\,
			T\gtrsim T_{\scriptscriptstyle{\mathrm{CBKT}}}\,.
        \label{eq:highTcurrent}
    \end{equation}
Notably, Eq.\,(\ref{eq:highTcurrent}) looks like a formula for thermally activated
conductivity.  Yet, one has to remember that the physical mechanism behind
the considered charge transfer is quantum mechanical tunnelling process which can take
place only if the mechanisms for the energy relaxation are switched on.

Turning now to very low temperatures, 
$T\ll T_{\scriptscriptstyle{\mathrm{CBKT}}}\simeq {E}_{\mathrm C}$,
one sees that from Eq.\,(\ref{eq:W}) follows that the characteristic energy
$W\simeq {E}_{\mathrm C}\exp(-\bar{E}_{\mathrm C}/T)$, 
i.e. the relaxation rate becomes exponentially low.
The unbound charges that have to mediate the energy relaxation from the
tunnelling carriers are in an exponentially short supply, 
and the estimate for conductivity yields:
    \begin{equation}
        \sigma\sim\exp[-\ln(L/a)\exp({E}_{\mathrm C}/T)]\,, \,\,\,T\ll 
				T_{\scriptscriptstyle{\mathrm{CBKT}}}\,,
        \label{eq:lowTcurrent}
    \end{equation}
in accordance with the earlier estimate~\cite{FVB}.   
One has to bear in mind though that the latter formula has indeed a character of a very crude estimate
showing that conductivity at $T<T_{\scriptscriptstyle{\mathrm{CBKT}}}$ is practically zero
since the spectrum of the environment excitations acquires the gap $\simeq E_{\mathrm{C}}$.  
In reality, this "zero-conductivity" regime 
will be shunted at very low temperatures by phonons and quantum fluctuations 
(the discussion of which we leave for the forthcoming publication) which will yield the finite conductivity.

Let us note that interpolation formula (\ref{eq:W})  becomes very inaccurate in the close 
vicinity of the transition, since it does not take into account the fact that 
the typical distance 
between the free unbound charges (called the correlation lenght,
 $\xi_{\scriptscriptstyle{\mathrm{CBKT}}}$) 
diverges upon approach the transition as~\cite{Minnhagen,2DJJA_Review}:
	\begin{equation}\label{corrlength}
		\xi_{\scriptscriptstyle{\mathrm{CBKT}}}\simeq (1/a^2)
		\exp\sqrt{\frac{b}{(T/T_{\scriptscriptstyle{\mathrm{CBKT}}})-1}}\,,
	\end{equation}
where $b$ is a constant of the order of unity.  
This implies that the density of the environment excitations
starts to drop exponentially $\propto \xi_{\scriptscriptstyle{\mathrm{CBKT}}}^{-2}$ 
and that one has to observe
an appreciably increasing and deviating from its Arrhenius form resistance 
as a precursor of the charge-BKT when approaching
this transition from the above.  
This should be seen as an upturn in the $\ln R$ vs. $1/T$ curves and one can attribute 
the hyperactivated behaviour found~\cite{TiNHA} in the 5 nm thin
disordered titanium nitride (TiN) film, which was, by its degree of disorder,
in the extreme proximity to the disorder-driven SIT, approaching it from
the insulating side.
To test it we took $\ln R_{\square}(T)$ vs. $1/T$ plots
from~\cite{TiNHA} for zero field and $B=0.3$\,T and fitted it with 
	\begin{equation}\label{eq:RBKT}
		R=R_0\exp\left( A\exp\sqrt{\frac{b}{(T/T_{\scriptscriptstyle{\mathrm{CBKT}}})-1}} \right)\,,
		\label{upturn}
	\end{equation}
following from the fact that at $T\approx T_{\scriptscriptstyle{\mathrm{CBKT}}}$ should be
$W\propto\xi_{\scriptscriptstyle{\mathrm{CBKT}}}^{-2}$.  
One sees an excellent agreement of this formula (solid curves)
with the experimental data. The dashed line indicates the thermally activated behaviour, and one sees that 
the upturn smoothly transforms into the Arrhenius slope, 
this justifies our assumption that indeed the activated behaviour 
is observed in the temperature region where the density of free dipole excitations 
is low and the logarithmic interaction remains 
still unscreened. 
\begin{figure}[t]
\includegraphics[width=0.7\linewidth]{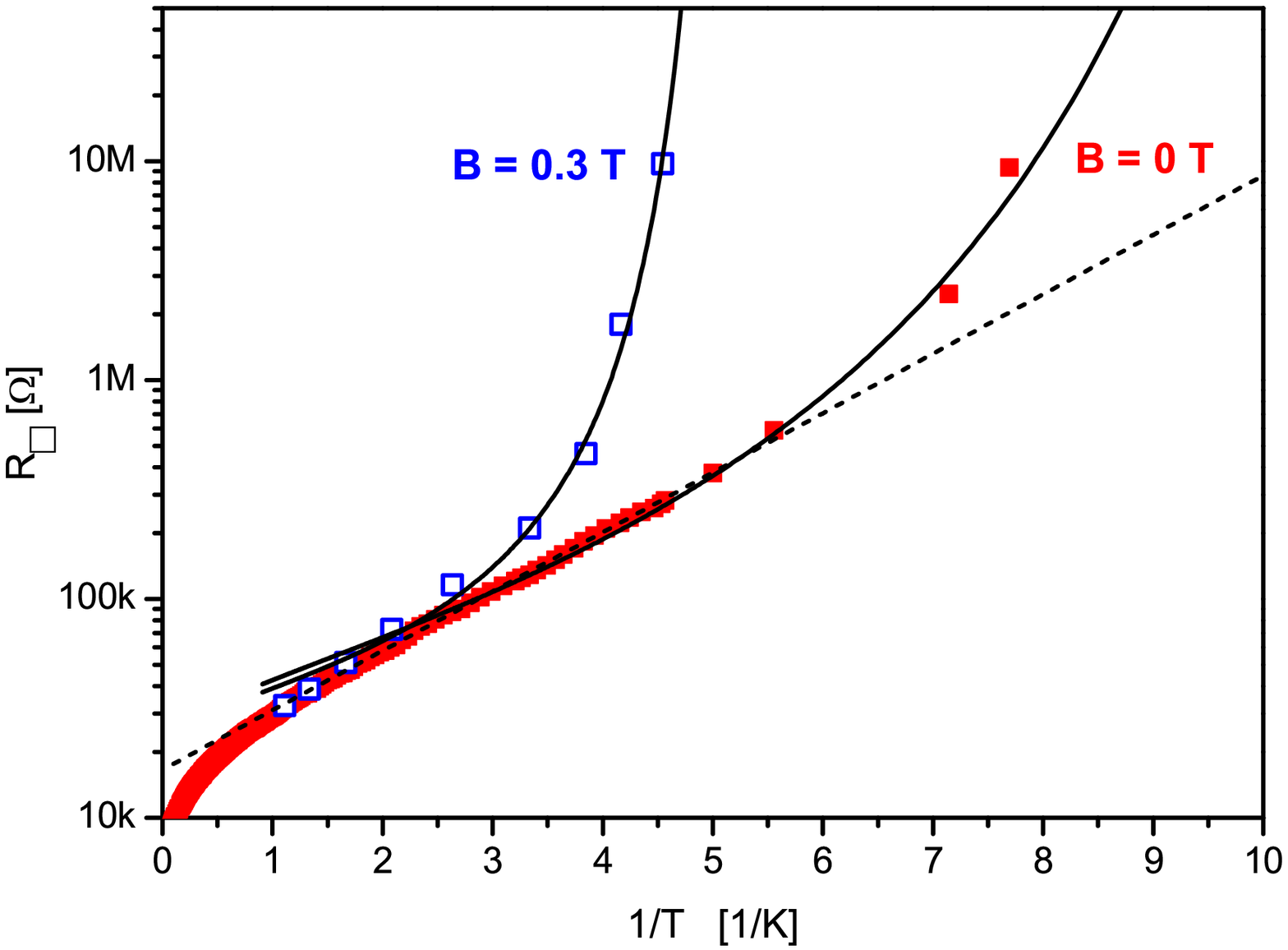}
\caption{ 
Plots of the logarithm of the sheet resistance $R_{\square}$ versus $1/T$ 
taken at two values of the  magnetic field, $B=0$\,T and $B=0.3$\,T
(data from ~\cite{TiNHA}) (red filled  and blue open symbols, respectfully). 
The solid lines represent the fit given by Eq.\,(\ref{eq:RBKT})
with parameters
 $T_{\scriptscriptstyle{\mathrm{CBKT}}}=0.062$\,mK, $b=1$, at $B=0$\,T, and 
 $T_{\scriptscriptstyle{\mathrm{CBKT}}}=0.175$\,mK, $b=0.5$ at $B=0.3$\,T. 
 The pre-exponential factors are the same for both curves and are taken as
$R_0 = 8$\,k$\Omega$ and $A=1$.  
The dashed straight line corresponds $T_I =0.63$\,K.
Equation\,(\ref{eq:RBKT}) is valid till the square root in the exponent exceeds unity. 
In the presented data this condition is satisfied at $T\leq 0.31$\,K for $B=0$ and
at $T\leq 0.35$\,K for $B=0.3$\,T. 
The overlap between the ranges of applicability of Eq.\,(\ref{eq:RBKT}) and activated behaviour means 
that indeed the screening length remains large enough down to 
$T\approx 1$\,K. Note that  $T_{\scriptscriptstyle{\mathrm{CBKT}}}=0.062$\,mK,
at zero field well  coincides with the above estimate (60\,mK) for a similar sample 
derived from the data on the size-dependent
activation energy.
}
\label{fig:FitTCBKT}
\end{figure}

As early as in `90-s,  an upturn in the $\log R(T)$ vs. $1/T$ dependence
indicating the ``stronger than activation" behavior was observed in 
JJAs~\cite{KandaCBKT,KandaChargeSoliton,Japan1}.  
They also viewed it as a precursor of the charge-BKT behaviour
and fitted it with 
$R\propto\exp\left[ A\exp\sqrt{{b}/[{(T/T_{\scriptscriptstyle{\mathrm{CBKT}}}])-1}}\right]$ 
analogously to the procedure adopted for description of the vortex-BKT data~\cite{2DJJA_Review}. 
We have found that such a procedure also gives a reasonable fit to the hyperactivated 
behaviour data of Ref.~\cite{TiNHA}, with slightly different fitting parameters.
This is not surprising since we work in the temperature range which 
does not include the extreme proximity to
$T_{\scriptscriptstyle{\mathrm{CBKT}}}$, 
i.e. in the range where $\xi_{\scriptscriptstyle{\mathrm{CBKT}}}$ is not excessively large.
 
\section{Applications}

\begin{figure}[t]
\includegraphics[width=\linewidth]{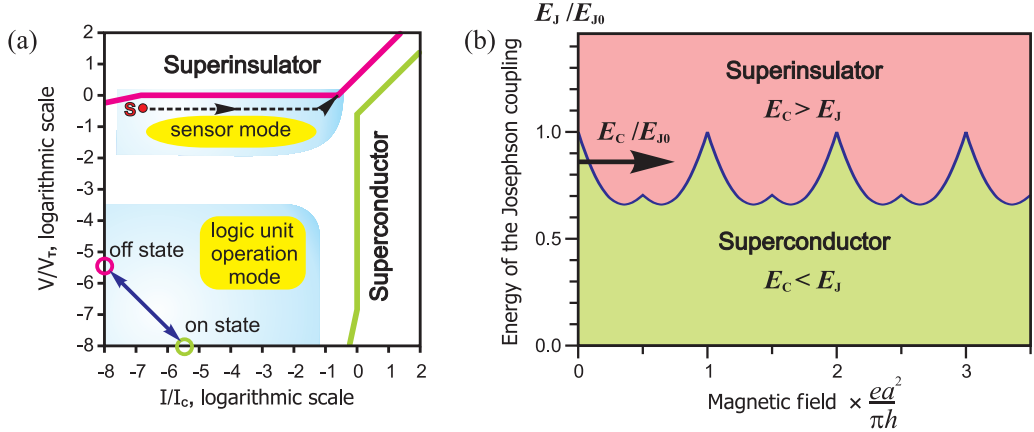}
\caption{
Switching between superconducting and superinsulating states. 
(a) Exemplary dual threshold current-voltage ($I$-$V$) 
characteristics in the superconducting and superinsulating states 
of a 5 nm thin TiN film in the double-log coordinates.  
The current and voltage are measured in units of the critical current, 
$I_\mathrm{c}$, and the threshold voltage, $V_\mathrm{T}$, respectively.  
The lower inset illustrates switching between the `off' (superinsulating) 
and `on' (superconducting) states in the logic unit operation mode.  
The upper inset illustrates the sensor (bolometer) mode 
utilizing the threshold character of the $I$-$V$ characteristics 
in the superinsulating state: when some pre-threshold voltage 
is applied to the film it remains superinsulating 
state with the zero current.  
As soon as the film is heated by some irradiation, 
the threshold voltage decreases below the applied value 
and the current jump over the six orders of magnitude occurs.  
An advantage of using a superinsulator as the working body 
for a sensor over the superconductor-based sensor is that 
due to huge resistance the current in the superinsulating state 
is extremely low, and, therefore, superinsulator 
is less sensitive to background noise. 
(b) A sketch of the magnetic filed controlled switching 
between two non-dissipative states 
in a square Josephson junction array (JJA).  
The line separating the superinsulating and superconducting 
domains depicts the dependence of Josephson coupling energy, 
$E_\mathrm{J}$ (measured in the units of the Josephson 
coupling energy, $E_\mathrm{J0}$, in the zero magnetic field), 
on the applied magnetic field, $B$.  
The arrow shows the evolution of the system, with the ratio 
$E_\mathrm{C}/E_\mathrm{J0}$ taken slightly less then unity, 
where $E_\mathrm{C}$ is the charging energy of a single junction.  
If $E_\mathrm{C}/E_\mathrm{J0} < 1$, the system is a superconductor, 
and if $E_\mathrm{C}/E_\mathrm{J0} > 1$, the system is superinsulator 
in the zero magnetic field.  
Upon increasing field, the system evolves along the arrow and, 
starting from the superconducting state transits into the superinsulator 
as soon as it crosses the line $E_\mathrm{J}(B)$, 
i.e. as soon as $E_\mathrm{J}(B)$ becomes less than 
the charging energy $E_\mathrm{C}$.
}
\label{fig:VAch}
\end{figure}

The existence of the superinsulating state opens
a route for a new class of devices for cryogenic electronics,
utilizing the duality between the superinsulation (SI) and superconductivity (SC). 
The left panel of Fig.\,\ref{fig:VAch} sketches the dual-diode current-voltage ($I$-$V$) characteristics
of a film or Josephson junction array 
corresponding to superconducting and superinsulating states respectively.
Application of a moderate voltage $V$ below the threshold
voltage $V_\mathrm{T}$ realizes the 'logical unit' operational mode of the device.
When in a superconducting state ('on' regime),
the device carries a loss-free supercurrent, in the superinsulating state
('off' regime) the current is blocked since $V < V_\mathrm{T}$.
Building the superswitch into an electric circuit, one designs
a binary logical unit, where (0) corresponds to, say,
the SI zero-current and (1) corresponds to the current-carrying SC state.
Note that in both, `on' and `off' regimes the superswitch
does not lose any power, remaining non-dissipative in both operating states.

\begin{figure}[t]
\includegraphics[width=\linewidth]{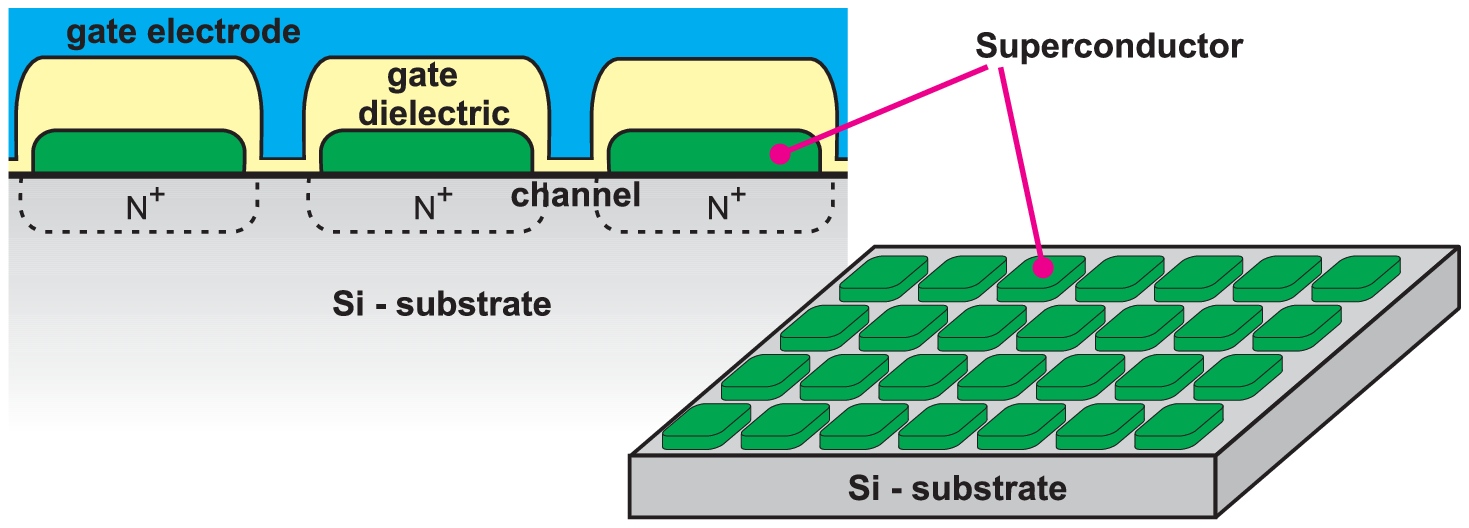}
\caption{
Superswitch utilizing the field effect transistor technology. 
A sketch of the cross section (the left panel) 
and the view of the working body (the right panel) 
of the superconductor-superinsulator field effect transistor (SSFET).  
The SSFET is comprised of the array of the superconducting 
islands placed on the Si-substrate.  
Highly doped regions of the substrate marked as $\mathrm{N}^+$.  
The islands are covered by a dielectric separating them from the gate on top. 
}
\label{fig:SSFET}
\end{figure}

Superinsulators can serve as a working body of
a detector or bolometer (see Fig.\,\ref{fig:VAch}).
To realize this mode one applies the bias $V$ less
but close to $V_\mathrm{T}$ and takes $T \lesssim T_\mathrm{SI}$ as a working temperature.
Heating by an irradiation shifts $V_\mathrm{T}$ below the voltage
of the working point and the current jump over several orders of magnitude occurs.
The choice of $T \lesssim T_\mathrm{SI}$ as a working point ensures
that the current jump is controlled by the heating instability solely,
eliminating effects of disorder and imperfections,
which would randomly shifted the voltage at which the switching
to the 'hot' branch takes place if the working point were chosen
deep in the superinsulating state.
The important feature of the prospective superinsulator-based
electronic devices is that one can expect their sensitivity 
to be very high since the current in the 
closed mode is completely blocked.  

The switching between the regimes, i.e. between the superconducting
and superinsulating states, can be implemented with the aid of
the magnetic field which modulates the Josephson coupling $E_\mathrm{J}$,
changing thus the actual $E_\mathrm{J}/E_\mathrm{C}$ ratio.  The Fig.\,\ref{fig:VAch}(b)
displays the energy-magnetic field phase diagram for a particular
case of the square JJA array, the $E_\mathrm{J}(B)$ dependence for which
was calculated in Ref.~\cite{Tinkham} in the nearest neighbors approximation.

From the point of view of applications the most appealing device would be the one that 
could be operated by means of the electric field. 
The technologically attractive possibility for implementation of such a device 
is to construct a planar Josephson junction array in which 
the electric field tunes the ratio $E_\mathrm{J}/E_{\mathrm C}$, and, therefore,
switches the device between the SI and SC states. 
A practical design of such a superswitch which is  realized by utilizing 
the field effect transistor principle, a 
superconductor-superinsulator field effect transistor (SSFET)
is shown in Fig.\,\ref{fig:SSFET}.
In this device in the absence of voltage at the gate the superconducting islands 
are decoupled due to high tunnelling resistance, $R_T$, of the dielectric separating them,
and the array is in a superinsulating state.  
Applying the sufficient gate voltage the insulating channels 
can be turned conducting (or, at least, their $R_T$ becomes reduced). 
This will give rise to increase of the ratio $E_\mathrm{J}/E_{\mathrm C}$ 
and will drive the array in a  superconducting state.  
Thus, the proposed switching technique is built on changing the 
conductivity of the medium confined between the adjacent superconducting islands 
by tuning the gate voltage. 

The first steps on the realization of the electrostatically-driven SIT have already been undertaken.
The system of the superconducting islands placed on the graphene layer was used 
in the recent work~\cite{ESITnmat2012} in order to tune the 
$E_\mathrm{J}/E_\mathrm{C}$ ratio as proposed above and the
pronounced SIT with the resistance as high as $10^7\,\Omega$ at 40\,mK 
was observed in the insulating state.
The change of the resistance as a function of the gate voltage was about seven orders of magnitude.

The story, however, would have been incomplete, without mentioning 
the approach based on the electric field-induced 
change of the properties of the superconducting films themselves~\cite{ReviewElectrostatics}.  
Such an approach, a direct varying of the charge carrier density 
in the superconducting film by applying a gate voltage in a field effect transistor 
configuration was taken in many works~\cite{GoldmanESIT,Matthey2007,Caviglia2008,Ueno2008,Bell2009,Biscaras2010,Ye2010,
Dhoot2010,suFET}.
At present two kinds of materials are used as gate insulators: 
SrTiO$_3$ (STO) crystal and/or various electrolytes (or ionic liquids).  
We would like to note that 
STO is an insulator with the very high dielectric constant $\varepsilon\sim 20 000$.
Such a dielectric sandwiching or covering the superconducting film is detrimental to formation 
of the superinsulating state
since it catastrophically reduces the electrostatic screening length, see Eq.\,(\ref{eq:Lambda}).
Indeed, in none of the quoted works 
the sheet resistance in the non-superconducting state does not exceed $40$\,K$\Omega$, thus the 
very possibility of the transition into an insulating state remains to be investigated.
Turning to the electrolytes as gate insulators, their disadvantage, 
from the viewpoint of the practical applications, is 
that the operating mode requires the thermocycling up to room temperatures for changing the 
gate voltage and influencing the properties of superconducting films~\cite{suFET}.
Yet, both novel techniques lend the way for controlling properties of superconducting films 
and open a route  to new generation switching devices with the outstanding 
performance characteristics.

\section{Results}

Our results can be summarized as follows:

\begin{enumerate}

\item On the fundamental level, the phenomenon of superinsulation rests on the duality 
between superconducting vortices and charges 
 in two-dimensional systems experiencing the superconductor-insulator transition.
 At the superconducting side logarithmic interaction between vortices (i.e. vortex confinement)
 leads to superconductivity, the state with the zero linear resistance below the vortex-BKT 
 transition where global phase coherence establishes.
 Correspondingly, the logarithmic interaction between charges (charge confinement) 
 at the insulating side leads to a superinsulating state
 with the zero linear conductivity. 
 This duality follows from the Heisenberg uncertainty principle: at the superconducting side 
 the phase uncertainty is zero implying that charges that comprise Cooper pair condensate 
 move without scattering.
 At the superinsulating side all the charges are pinned below the charge-BKT 
 transition implying, according to the Aharonov-Casher effect, the
 coherent quantum tunnelling of all the fluxons (the synchronized quantum slips 
 of globally uncertain phase) i.e. that fluxons are in a superfluid state. 

\item The condition for a charge confinement i.e. logarithmic growth of 
the interaction between the charges 
as a function of the distance between them in
the insulating state of a planar Josephson junction array is realized if the electrostatic 
screening length exceeds the linear size of the array.
This requires that a single junction capacitance well exceeds the capacitance 
to the ground.   
This logarithmic interaction leads to the logarithmic scaling 
of the activation energy of the Cooper pair insulator in Josephson junctions array 
with the linear size of the array. 
We refer to this phenomenon of emergence of the large energy scale 
controlling the activated behaviour and
growing with the size of the system as to a \textit{macroscopic Coulomb blockade}. 

\item Disordered thin films in the critical vicinity of the superconductor-superinsulator transition have 
huge, diverging on approach to the transition to the superconducting state, dielectric constant.  
This leads to the two-dimensional logarithmic Coulomb interaction 
between the unscreened charges in the film,
which, in its turn results in the charge Berezinskii-Kosterlitz-Thouless transition.  
The logarithmic interaction holds as long as 
the distance separating charges does not exceed the electrostatic screening length.
Accordingly the CBKT transition is most
pronounced in the systems whose linear size does not exceed this length.

\item One can expect that in the critical vicinity of the superconductor-superinsulator transition 
where long-range Coulomb interactions become relevant, electronic phase separation can 
occur and the texture of weakly coupled superconducting islands may form.  
This indicates that near the transition the transport behaviour 
of disordered films is analogous to that of planar Josephson junctions arrays.

\item Logarithmic interaction between charges in the Cooper-pair insulating phase of the
two-dimensional Josephson junction array 
is dual to the  logarithmic interactions between the vortices at the superconducting side.  
This implies the dual phase diagram in the vicinity of the superconductor-superinsulator transition:
Transformation of the resistive state into superconductivity possessing the global phase coherence
 occurring at $T_{\mathrm{\scriptscriptstyle{VBKT}}}$ is mirrored by the transition 
from the insulator to superinsulator at the charge BKT transition temperature 
$T_{\mathrm{\scriptscriptstyle{CBKT}}}$.  
We identify a superinsulator as a low-temperature charge-BKT phase, 
possessing infinite resistivity in the finite temperature range in the 
same sense as the low-temperature vortex BKT phase is
a superconductor having infinite conductivity in a finite temperature range as well.

\item In the Cooper-pair insulator the current occurs via tunnelling of the Cooper pairs 
across the Josephson links.  
To ensure tunnelling transport in the random system, the energy relaxation 
mechanism providing the energy exchange between the charge carriers and some 
bosonic environment is required.  
At low temperatures tunnelling in Josephson junction array is mediated by the
\textit{self-generated} bosonic environment 
of the Cooper-pair dipole excitations, comprised of the same Cooper pairs
that tunnel and mediate the charge transfer.   
Importantly, being a system with the infinite number degrees of freedom, 
the dipole environment plays itself the role of the thermostat.  
Opening the gap in the \textit{local} dipole excitations spectrum suppresses 
energy relaxation and impedes the tunnelling current.  
In the two-dimensional Josephson junction array  this gap appears below 
$T_{\mathrm{\scriptscriptstyle{CBKT}}}$.
We expect that the gap in the local dipole spectrum is associated 
with formation of a low-temperature glassy  state. 
This constitutes the microscopic mechanism of superinsulation. 
 
\end{enumerate}

\section{Conclusions}

In our work we have shown that the existence of the superinsulating state 
is a fundamental phenomenon intimately connected with the very existence
of superconductivity and that it follows from the symmetry of the uncertainty principle 
setting a competition between the conjugated quantities, 
the charge and the phase of the wave function of the Cooper pair condensate.  
This symmetry can be expressed in terms of the charge-vortex duality in two-dimensional 
superconducting systems and manifests itself via the reversal between 
the Aharonov-Bohm and the Aharonov-Casher effects.
We have described the conditions for emerging of superinsulation in real experimentally 
accessible two-dimensional systems,
lateral Josephson junction arrays and highly disordered thin superconducting films 
and demonstrated that the critical component of the phenomenon
is the high dielectric constant of the insulating phase of the system in question 
which develops in the close vicinity of superconductor-insulator transition.
The latter ensures the logarithmic interaction between the charges and brings with it 
the effect of the macroscopic Coulomb blockade, which, in its turn
manifests as a size-dependent characteristic energy that controls activation processes 
in the insulating phase. 
We examined these concepts testing them on the original experimental findings 
and demonstrated that they work perfectly well offering self consistent physical picture
describing collective behaviour of two-dimensional superconducting systems in the vicinity 
of superconductor-insulator transition.   
We constructed and analysed the phase diagram for a planar Josephson junction array 
and/or disordered superconducting film in the vicinity of 
the superconductor insulator transition and identified superinsulating state 
as a low temperature charge-BKT phase.  
And, finally, we discussed the perspectives for practical applications of the superinsulating systems 
as a working body for new generation of electronic devices
utilizing the duality between superinsulation and superconductivity.

\section*{Acknowledgements}

We are delighted to thank D.\,Averin, N.\,Chtchelkatchev, E.\,Chudnovsky,
R.\,Fazio, A.\,Glatz, L.\,Glazman, K.\,Matveev, A.\,Melnikov, V.\,Mineev,  Yu.\,Nazarov, 
and Ch. Strunk for enlightening discussions.
We are most grateful  to D.\,Khomskii for careful reading of the manuscript and useful suggestions.
The work was supported by the U.S. Department of Energy Office of Science through
the contract DE-AC02-06CH11357.
The work of TB was partly supported by the Program
``Quantum Mesoscopic and Disordered Systems" of the Russian Academy of Sciences
and by the Russian Foundation for Basic Research (Grant No. 12-02-00152).


\begin{thebibliography}{00}
\bibitem{Anderson64} 
P.~W. Anderson,  Lectures on the Many-Body Problem, Vol.
2, edited by E.~R. Caianiello (Academic, New York 1964) 113.

\bibitem{Abeles77}
B.~Abeles,
Effect of charging energy on superconductivity in granular metal films,
Phys. Rev. B 15 (1977) 2828--2829.

\bibitem{Efetov1980} 
K.~B.~Efetov, 
Phase transitions in granulated superconductors,
Zh. Eksp. Teor. Fiz. 78 (1980) 2017--2032, 
[Sov. Phys. JETP. 51 (1980) 1015--1022].

\bibitem{Gold1983ZPhys} 
A.~Gold, 
Impurity-Induced Phase Transition in the Interacting Bose Gas: 1. Analytical Results,
Z. Phys. B - Condensed Matter 52 (1983) 1--8.

\bibitem{Gold1986PRA} 
A.~Gold, 
Dielectric properties of a disordered Bose condensate, 
Phys. Rev. A 33 (1986) 652–-659.

\bibitem{Feynman}  
R.~D. Feynman, Statistical Mechanics: A Set of Lectures, Addison Wesley, 1981.

\bibitem{Sugahara85} 
M.~Sugahara, 
Superconductive Granular Thin Film and Phase Quantum Tunnel Device,
Jpn. J. Appl. Phys. 24 (1985) 674--678.

\bibitem{Mooij2006} 
J.~E.~Mooij and Yu.~V.~Nazarov,
Superconducting nanowires as quantum phase-slip junctions,
Nature Physics 2 (2006) 169--172. 

\bibitem{Beasley1979} 
M.~R.~Beasley,  J.~E.~Mooij,  and T.~P.~Orlando, 
Possibility of Vortex-Antivortex Pair Dissociation in Two-Dimensional Superconductors, 
Phys. Rev. Lett. 42 (1979) 1165--1168.

\bibitem{HalperinNelson1979} 
B.~I.~Halperin  and D.~R.~Nelson, 
Resistive Transitions in Superconducting Films,
J. Low. Temp. Phys. 36 (1979) 599--616.

\bibitem{FVB}
M.~V.~Fistul, V.~M.~Vinokur, and  T.~I.~Baturina, 
Collective Cooper-Pair Transport in the Insulating State of Josephson-Junction Arrays,
Phys. Rev. Lett. 100 (2008) 086805.

\bibitem{VinNature} 
V.~M.~Vinokur, T.~I.~Baturina, M.~V.~Fistul, A.~Yu.~Mironov, M.~R.~Baklanov, and C.~Strunk,
Superinsulator and quantum synchronization,
Nature 452 (2008) 613--615.

\bibitem{Minnhagen} 
P.~Minnhagen, 
The two-dimensional Coulomb gas, vortex unbinding, and superfluid-superconducting films,
Rev. Mod. Phys. 59 (1987) 1001--1066.

\bibitem{Mooij1989} 
L.~J.~Geerligs, M.~Peters, L.~E.~M.~de~Groot, A.~Verbruggen, and J.~E.~Mooij, 
Charging effects and quantum coherence in regular Josephson junction arrays,
Phys. Rev. Lett. 63 (1989) 326--329.

\bibitem{Mooij1990} 
J.~E.~Mooij, B.~J.~van~Wees, L.~J.~Geerligs, M.~Peters, R.~Fazio, and G.~Sch\"on, 
Unbinding of charge-anticharge pairs in two-dimensional arrays of small tunnel junctions,
Phys. Rev. Lett. 65 (1990) 645--648.

\bibitem{FazioSchon1991} 
R.~Fazio and G.~Sch\"on,
Charge and vortex dynamics in arrays of tunnel junctions,
Phys. Rev. B 43 (1991) 5307--5320.

\bibitem{WeesDual} 
B.~J.~van~Wees,
Duality between Cooper-pair and vortex dynamics in two-dimensional Josephson-junction arrays,
Phys. Rev. B 44 (1991) 2264--2267.

\bibitem{Mooij1992} 
H.~S.~J.~van~der~Zant, F.~C.~Fritschy, W.~J.~Elion, L.~J.~Geerligs, and J.~E.~Mooij, 
Field-induced superconductor-to-insulator transitions in Josephson-junction arrays,
Phys. Rev. Lett. 69 (1992) 2971--2974.

\bibitem{Tighe} 
T.~S.~Tighe, M.~T.~Tuominen, J.~M.~Hergenrother, and M.~Tinknam, 
Measurements of charge soliton motion in two-dimensional arrays of ultrasmall Josephson junctions,
Phys. Rev. B 47 (1993) 1145--1148.

\bibitem{Haviland1} 
P.~Delsing, C.~D.~Chen, D.~B.~Haviland, Y.~Harada, and T.~Claeson, 
Charge solitons and quantum fluctuations in two-dimensional arrays of small Josephson junctions,
Phys. Rev. B 50 (1994) 3959--3971. 

\bibitem{KandaChargeSoliton} 
A.~Kanda, S.~Katsumoto, and S.~Kobayashi, 
Charge-Soliton Transport Properties in Two-Dimensional Array of Small Josephson Junctions, 
J. Phys. Soc. Jpn. 63 (1994) 4306--4309. 

\bibitem{KandaCBKT} 
A.~Kanda and S.~Kobayashi, 
Precursor of Charge KTB Transition in Normal and Superconducting 
Tunnel Junction Array, 
J. Phys. Soc. Jpn. 64 (1995) 19--21. 

\bibitem{KandaSelfcapacitance} 
A.~Kanda and S.~Kobayashi, 
Effect of Self-Capacitance on Charge Kosterlitz-Thouless Transition in Small Tunnel-Junction Arrays, 
J. Phys. Soc. Jpn. 64 (1995) 3172--3174. 

\bibitem{Mooij1996} 
H.\,S.\,J.~van~der~Zant, W.~J.~Elion, L.\,J.~Geerligs, and J.\,E.~Mooij, 
Quantum phase transitions in two dimensions: Experiments in Josephson-junction arrays,
Phys. Rev. B 54 (1996) 10081--10093.

\bibitem{Japan1} 
T.~Yamaguchi, R.~Yagi, S.~Kobayashi, and Y.~Ootuka,
Two-Dimensional Arrays of Small Josephson Junctions with Regular and Random Defects,
J. Phys. Soc. Jpn. 67 (1998) 729--731.

\bibitem{Japan2} 
Y.~Takahide, R.~Yagi, A.~Kanda, Y.~Ootuka, and S.~Kobayashi, 
Superconductor-Insulator Transition in a Two-Dimensional Array 
of Resistively Shunted Small Josephson Junctions,
Phys. Rev. Lett. 85 (2000) 1974--1977.

\bibitem{Review1} 
Macroscopic Quantum Phenomena and
Coherence in Superconducting Networks, edited by C. Giovanella and
M. Tinkham, World Scientific, Singapore, 1995.

\bibitem{2DJJA_Review}
R.\,S. Newrock, C.\,J. Lobb, U. Geigenm\"uller, and M. Octavio,
The Two-Dimensional Physics of Josephson Junction Arrays,
Solid State Phys. 54 (2000) 263--512.

\bibitem{FazioZant2001} 
R.~Fazio and H.~van~der~Zant, 
Quantum phase transitions and vortex dynamics in superconducting networks,
Physics Reports 355 (2001) 235--334.

\bibitem{Dynes78} 
R.~C.~Dynes, J.~P.~Garno, and J.~M.~Rowell,
Two-Dimensional Electrical Conductivity in Quench-Condensed Metal Films,
Phys. Rev. Lett. 40 (1978) 479--482.

\bibitem{White} 
Alice~E.~White, R.~C.~Dynes, and J.~P.~Garno,
Destruction of superconductivity in quench-condensed two-dimensional films,
Phys. Rev. B 33 (1986) 3549--3552.

\bibitem{Orr86} 
B.~G.~Orr, H.~M.~Jaeger, A.~M.~Goldman, and C.~G.~Kuper, 
Global Phase Coherence in Two-Dimensional Granular Superconductors,
Phys. Rev. Lett. 56 (1986) 378--381.

\bibitem{SugaharaLT18} 
N.~Yoshikawa, T.~Akeyoshi, M.~Kojima, and M.~Sugahara,
Dual Conduction Characteristics Observed in Highly Resistive
NbN Granular Thin Films,
Proc. LT-18, Kyoto 1987, Jpn. J. Appl. Phys. 26 (Suppl.
26-3) (1987) 949--950.

\bibitem{Sugahara26L} 
N.~Yoshikawa, T.~Akeyoshi, and M.~Sugahara,
Field-Effect Induced Sinusoidal Conductivity Variation of NbN Granular Thin Films,
Jpn. J. Appl. Phys. 26 (1987) L1701--L1702.

\bibitem{Widom}
A.~Widom and S.~Badjou,
Quantum displacement-charge transitions in two-dimensional granular superconductors, 
Phys. Rev. B 37 (1988) 7915--7916.

\bibitem{Jaeger} 
H.~M.~Jaeger, D.~B.~Haviland, B.~G.~Orr, and A.~M. Goldman,
Onset of superconductivity in ultrathin granular metal films,
Phys. Rev. B 40 (1989) 182--196.

\bibitem{Barber90} 
R.~P.~Barber, Jr. and R.~E.~Glover III,
Hyper-resistivity to global-superconductivity transition by annealing in quench-condensed Pb film,
Phys. Rev. B 42 (1990) 6754--6757.

\bibitem{Wu94} 
W.~Wu and P.~W.~Adams,
Electric-field tuning of the superconductor-insulator transition in granular Al films,
Phys. Rev. B 50 (1994) 13065--13068.

\bibitem{GoldmanGranular} 
C.~Christiansen, L.~M.~Hernandez, and A.~M.~Goldman,
Evidence of Collective Charge Behavior in the Insulating State 
of Ultrathin Films of Superconducting Metals,
Phys. Rev. Lett. 88 (2002) 037004.

\bibitem{Frydman}
A.~Frydman,
The superconductor insulator transition in systems of ultrasmall grains,
Physica C 391 (2003) 189--195.

\bibitem{Barber06} 
R.~P.~Barber, Jr., Shih-Ying~Hsu, J.~M.~Valles, Jr., R.~C.~Dynes, and R.~E.~Glover III,
Negative magnetoresistance, negative electroresistance, and metallic behavior 
on the insulating side of the two-dimensional superconductor-insulator transition in granular Pb films,
Phys. Rev. B 73 (2006) 134516.

\bibitem{Strongin70} 
M.~Strongin, R.~S.~Thompson, O.~F.~Kammerer, and J.~E.~Crow, 
Destruction of Superconductivity in Disordered Near-Monolayer Films, 
Phys. Rev. B 1 (1970) 1078--1091.

\bibitem{HavGold} 
D.~B.~Haviland, Y.~Liu, and A.~M.~Goldman,
Onset of Superconductivity in the Two-Dimensional Limit,
Phys. Rev. Lett. 62 (1989) 2180--2183.

\bibitem{Hebard} 
A.~F.~Hebard and M.~A.~Paalanen, 
Magnetic-field-tuned superconductor-insulator transition in two-dimensional films,
Phys. Rev. Lett. 65 (1990) 927--930.

\bibitem{ShaharOvadyahu} 
D.~Shahar and Z.~Ovadyahu,
Superconductivity near the mobility edge,
Phys. Rev. B 46 (1992) 10917--10922.

\bibitem{LiuGoldman} 
Y.~Liu, D.~B.~Haviland, B.~Nease, and A.~M.~Goldman,
Insulator-to-superconductor transition in ultrathin films,
Phys. Rev. B 47 (1993) 5931--5946.

\bibitem{Zvi1994} 
D.~Kowal, Z.~Ovadyahu,
Disorder induced granularity in an amorphous superconductor,
Solid St. Comm. 90 (1994) 783--786.

\bibitem{VFGIns} 
V.~F.~Gantmakher, M.~V.~Golubkov, J.~G.~S.~Lok, and A.~K.~Geim,
Giant negative magnetoresistance of semi-insulating amorphous indium oxide 
films in strong magnetic fields,
Zh. Eksp. Teor. Fiz. 109 (1996) 1765--1778 
[JETP 82 (1996) 951--958].

\bibitem{Marcovic98} 
N.~Markovi\'{c}, C.~Christiansen, and A.~M.~Goldman, 
Thickness-Magnetic Field Phase Diagram at the Superconductor-Insulator Transition in 2D,
Phys. Rev. Lett. 81 (1998) 5217--5220. 

\bibitem{VFG1998} 
V.~F.~Gantmakher, M.~V.~Golubkov, V.~T.~Dolgopolov, G.~E.~Tsydynzhapov, and A.~A.~Shashkin, 
Destruction of localized electron pairs above the magnetic-field-driven superconductor-insulator
transition in amorphous In-O films, 
JETP Lett. 68 (1998) 363--369.

\bibitem{Valles1999} 
J.~A.~Chervenak and J.~M.~Valles, Jr., 
Observation of critical amplitude fluctuations near the two-dimensional 
superconductor-insulator transition, 
Phys. Rev. B 59 (1999) 11209--11212.

\bibitem{VFGScaling} 
V.~F.~Gantmakher, M.~V.~Golubkov, V.~T.~Dolgopolov, G.~E.~Tsydynzhapov, and A.~A.~Shashkin, 
Scaling Analysis of the Magnetic Field-Tuned Quantum Transition 
in Superconducting Amorphous In-O Films, 
JETP Lett. 71 (2000) 160--164.

\bibitem{VFGPar2000}
V.~F.~Gantmakher, M.~V.~Golubkov, V.~T.~Dolgopolov, A.~A.~Shashkin, and G.~E.~Tsydynzhapov, 
Observation of the Parallel-Magnetic-Field-Induced Superconductor-Insulator Transition
in Thin Amorphous In-O Films, 
JETP Lett. 71 (2000) 473--478.

\bibitem{ButkoAdams} 
V.~Yu.~Butko and P.~W.~Adams, 
Quantum metallicity in a two-dimensional insulator,
Nature 409 (2001) 161--164.

\bibitem{Be1} 
E.~Bielejec, J.~Ruan, and Wenhao~Wu,
Hard Correlation Gap Observed in Quench-Condensed Ultrathin Beryllium,
Phys. Rev. Lett. 87 (2001) 036801.

\bibitem{BeInsPMR} 
E.~Bielejec, J.~Ruan, and Wenhao~Wu,
Anisotropic magnetoconductance in quench-condensed ultrathin beryllium films,
Phys. Rev. B 63 (2001) 100502(R).

\bibitem{BeBSIT} 
E.~Bielejec and Wenhao~Wu,
Field-Tuned Superconductor-Insulator Transition with and without Current Bias,
Phys. Rev. Lett. 88 (2002) 206802.

\bibitem{Shahar-act} 
G.~Sambandamurthy, L.~W.~Engel, A.~Johansson, and D.~Shahar, 
Superconductivity-Related Insulating Behavior,
Phys. Rev. Lett. 92 (2004) 107005.

\bibitem{TBJETPL} 
T.~I.~Baturina, D.~R.~Islamov, J.~Bentner, C.~Strunk, M.~R.~Baklanov, and A.~Satta,
Superconductivity on the Localization Threshold 
and Magnetic-Field-Tuned Superconductor-Insulator Transition in TiN Films,
Pis'ma Zh. Eksp. Teor. Fiz. 79 (2004) 416--420 
[JETP Lett. 79 (2004) 337-341].

\bibitem{Hadacek} 
N.~Hadacek, M.~Sanquer, and J-C.~Vill\'{e}gier, 
Double reentrant superconductor-insulator transition in thin TiN films,
Phys. Rev. B 69 (2004) 024505. 

\bibitem{TiNPhysB2005}
T.~I.~Baturina, J.~Bentner, C.~Strunk, M.~R.~Baklanov, A.~Satta, 
From quantum corrections to magnetic-field-tuned superconductor-insulator quantum phase
transition in TiN films, 
Physica B 359-361 (2005) 500--502.

\bibitem{SITInOKapitulnik}
Myles Steiner  and Aharon Kapitulnik,
Superconductivity in the insulating phase above the field-tuned superconductor-insulator transition
in disordered indium oxide films, 
Physica C 422 (2005) 16--26.

\bibitem{Shahar-Coll} 
G.~Sambandamurthy, L.~W.~Engel, A.~Johansson, E.~Peled, and D.~Shahar, 
Experimental Evidence for a Collective Insulating State in Two-Dimensional Superconductors,
Phys. Rev. Lett. 94 (2005) 017003.

\bibitem{Xiong2006}
J.~S.~Parker, D.~E.~Read, A.~Kumar, and P.~Xiong, 
Superconducting quantum phase transitions tuned by magnetic impurity and magnetic field 
in ultrathin a-Pb films,
Europhys. Lett. 75 (2006) 950--956.

\bibitem{TiNQM} 
T.~I.~Baturina, C.~Strunk, M.~R.~Baklanov, and A.~Satta,
Quantum Metallicity on the High-Field Side of the Superconductor-Insulator Transition,
Phys. Rev. Lett. 98 (2007) 127003.

\bibitem{TiNSIT} 
T.~I.~Baturina, A.~Yu.~Mironov, V.~M.~Vinokur, M.~R.~Baklanov, and C.~Strunk, 
Localized Superconductivity in the Quantum-Critical Region of the Disorder-Driven 
Superconductor-Insulator Transition in TiN Thin Films, 
Phys. Rev. Lett. 99 (2007) 257003.

\bibitem{TiNPhysC}
T.~I.~Baturina, A.~Bilu\v{s}i\'{c}, A.~Yu.~Mironov, V.~M.~Vinokur, M.~R.~Baklanov, and C.~Strunk, 
Quantum-critical region of the disorder-driven superconductor-insulator transition,
Physica C 468 (2008) 316--321.

\bibitem{Zvi2008} 
D.~Kowal and Z.~Ovadyahu, 
Scale dependent superconductor-insulator transition,
Physica C 468 (2008) 322--325. 

\bibitem{TiNHA} 
T.~I.~Baturina, A.~Yu.~Mironov, V.~M.~Vinokur, M.~R.~Baklanov, and C.~Strunk,
Hyperactivated Resistance in TiN Films on the Insulating Side of the Disorder-Driven
Superconductor-Insulator Transition, 
JETP Lett. 88 (2008) 752--757.

\bibitem{Goldman2010}
Yen-Hsiang~Lin and A.~M.~Goldman,
Hard energy gap in the insulating regime of nominally granular films near the 
superconductor-insulator transition,
Phys. Rev. B 82 (2010) 214511. 

\bibitem{ChargeBKT} 
D.~Kalok, A.~Bilu\v{s}i\'{c}, T.~I.~Baturina,  V.~M.~Vinokur, and C.~Strunk, 
Intrinsic non-linear conduction in the super-insulating state of thin TiN films,
arXiv:1004.5153v2 (2010).

\bibitem{Goldman2011}
Yen-Hsiang~Lin and A.~M.~Goldman, 
Magnetic-Field-Tuned Quantum Phase Transition in the Insulating Regime of Ultrathin
Amorphous Bi Films, 
Phys. Rev. Lett. 106 (2011) 127003.

\bibitem{ShaharAngular}
A.~Johansson, I.~Shammassa, N.~Stander, E.~Peled, G.~Sambandamurthy, D.~Shahar, 
Angular dependence of the magnetic-field driven superconductor-insulator 
transition in thin films of amorphous indium-oxide, 
Solid State Comm. 151 (2011) 743--746.

\bibitem{ShaharSteps}
O.~Cohen, M.~Ovadia, and D.~Shahar, 
Electric breakdown effect in the current-voltage characteristics of amorphous indium oxide thin
films near the superconductor-insulator transition, 
Phys. Rev. B 84 (2011) 100507(R). 

\bibitem{KalokLT26}
D.~Kalok, A.~Bilu\v{s}i\'{c}, T.~I.~Baturina, A.~Yu.~Mironov, S.~V.~Postolova, A.~K.~Gutakovskii, 
A.~V.~Latyshev, V.~M.~Vinokur, and C.~Strunk, 
Non-linear conduction in the critical region of the superconductor-insulator transition in TiN thin films, 
Journal of Physics: Conference Series (JPCS), accepted for publication.

\bibitem{Mandrus1991}
D.~Mandrus, L.~Forro, C.~Kendziora, and L.~Mihaly ,
Two-dimensional electron localization in bulk single crystals of 
Bi$_2$Sr$_2$Y$_x$Ca$_{1-x}$Cu$_2$O$_8$,
Phys. Rev. B 44 (1991) 2418–-2421.

\bibitem{Rosenbaum1992}
G.~T.~Seidler, T.~F.~Rosenbaum, B.~W.~Veal,
Two-dimensional superconductor-insulator transition in bulk single-crystal YBa$_2$Cu$_3$O$_{6.38}$,
Phys. Rev. B 45 (1992) 10162--10164.
 
\bibitem{Tanda1992}
Satoshi Tanda, Shigeki Ohzeki, and Tsuneyoshi Nakayama,
Bose glass-vortex–glass phase transition and dynamics scaling for high-$T_c$ 
Nd$_{2-x}$Ce$_x$CuO$_4$ thin films, 
Phys. Rev. Lett. 69 (1992) 530--533.

\bibitem{Beschoten1996}
B.~Beschoten, S.~Sadewasser, G.~G\"untherodt, and  C.~Quitmann,  
Coexistence of Superconductivity and Localization in Bi$_2$Sr$_2$(Ca$_z$, Pr$_{1-z}$)Cu$_2$O$_{8+y}$, 
Phys. Rev. Lett. 77 (1996) 1837--1840.  

\bibitem{LavrovSIT}
Yoichi Ando, A.~N.~Lavrov, Seiki Komiya, Kouji Segawa, and X.~F.~Sun, 
Mobility of the Doped Holes and the Antiferromagnetic Correlations in Underdoped High-$T_c$ Cuprates, 
Phys. Rev. Lett. 87 (2001) 017001. 
 
\bibitem{VFG_JETPL2003}
V.~F.~Gantmakher, S.~N.~Ermolov, G.~E.~Tsydynzhapov, A.~A.~Zhukov, and T.~I.~Baturina, 
Suppression of 2D Superconductivity by the Magnetic Field:
Quantum Corrections vs. the Superconductor-Insulator Transition, 
JETP Lett. 77 (2003) 424–-428. 

\bibitem{suFET}
A.~T.~Bollinger, G.~Dubuis, J.~Yoon, D.~Pavuna, J.~Misewich, and I.~Bo\v{z}ovi\'{c}, 
Superconductor-insulator transition in La$_{2-x}$Sr$_x$CuO$_4$ at the pair quantum resistance, 
Nature 472 (2011) 458--460.

\bibitem{LarkinOvchin71n1} 
A.~I.~Larkin and Yu.~N.~Ovchinnikov, 
Effect of inhomogeneities on the properties of superconductors, 
Zh. Eksp. Teor. Fiz. 61 (1971) 1221--1230 
[Sov. Phys. JETP 34 (1972) 651].

\bibitem{IoffeLarkin} 
L.~B.~Ioffe and A.~I.~Larkin, 
Properties of superconductors with a smeared transition temperature,
Zh. Eksp. Teor. Fiz. 81 (1981) 707--718  
[Sov. Phys. JETP 54 (1981) 378-384].

\bibitem{MaLee} 
Michael Ma and Patrick A. Lee,
Localized superconductors, 
Phys. Rev. B 32 (1985) 5658--5667.

\bibitem{Imry} 
Y.~Imry, M.~Strongin, and C.~C.~Homes,
An inhomogeneous Josephson phase in thin film and high-$T_c$ superconductors, 
Physica C 468 (2008) 288--293.

\bibitem{Feig2005}
M.~A.~Skvortsov and M.~V.~Feigel’man, 
Superconductivity in Disordered Thin Films: Giant Mesoscopic Fluctuations, 
Phys. Rev. Lett. 95 (2005) 057002.

\bibitem{GhosalPRL} 
A.~Ghosal, M.~Randeria, and N.~Trivedi,
Role of Spatial Amplitude Fluctuations in Highly Disordered s-Wave Superconductors,
Phys. Rev. Lett. 81 (1998) 3940--3943.

\bibitem{GhosalPRB} 
A.~Ghosal, M.~Randeria, and N.~Trivedi,
Inhomogeneous pairing in highly disordered s-wave superconductors,
Phys. Rev. B 65 (2001) 014501. 

\bibitem{DubiNat} 
Y.~Dubi, Y.~Meir, and Y.~Avishai, 
Nature of the superconductor-insulator transition in disordered superconductors, 
Nature 449 (2007) 876--880.

\bibitem{Trivedi2011} 
Karim Bouadim, Yen Lee Loh, Mohit Randeria and Nandini Trivedi, 
Single- and two-particle energy gaps across the disorder-driven superconductor-insulator transition, 
Nature Phys. 7 (2011) 884--889.

\bibitem{BaturinaSTM08} 
B.~Sac\'{e}p\'{e}, C.~Chapelier, T.~I.~Baturina, V.~M.~Vinokur,  M.~R.~Baklanov, and M.~Sanquer, 
Disorder-Induced Inhomogeneities of the Superconducting State
Close to the Superconductor-Insulator Transition, 
Phys. Rev. Lett. 101 (2008) 157006.

\bibitem{InOSTM_NP}
B.~Sac\'{e}p\'{e}, T.~Dubouchet, C.~Chapelier, M.~Sanquer, M.~Ovadia, D.~Shahar, M.~Feigel'man, 
and L.~Ioffe, 
Localization of preformed Cooper pairs in disordered superconductors,
Nature Phys. 7 (2011) 239--244. 

\bibitem{KapitulnikSTM}
A.~C.~Fang, L.~Capriotti, D.~J.~Scalapino, S.~A.~Kivelson, N.~Kaneko, 
M.~Greven, and A.~Kapitulnik, 
Gap-Inhomogeneity-Induced Electronic States in Superconducting 
Bi$_2$Sr$_2$CaCu$_2$O$_{8-\delta}$, 
Phys. Rev. Lett. 96 (2006) 017007.

\bibitem{YazdaniSTM}
K.~K.~Gomes, A.~N.~Pasupathy, A.~Pushp, Sh.~Ono, Y.~Ando, and A.~Yazdani, 
Visualizing pair formation on the atomic scale in the
high-$T_c$ superconductor 
Bi$_2$Sr$_2$CaCu$_2$O$_{8+\delta}$, 
Nature 447 (2007) 569--572.

\bibitem{FazioNatNV} 
R.~Fazio, 
Condensed-matter physics - Opposite of a superconductor, 
Nature 452 (2008) 542-543.

\bibitem{ArBohm}
Y. Aharonov and D. Bohm,
Significance of Electromagnetic Potentials in the Quantum Theory,
Phys. Rev. 115 (1959) 485-491.

\bibitem{ArCash}
Y. Aharonov  and A. Casher,
Topological Quantum Effects for Neutral Particles,
Phys. Rev. Lett. 53 (1984) 319-321.

\bibitem{RezAr} 
B. Reznik and Y. Aharonov,
Question of the nonlocality of the Aharonov-Casher effect,
Phys. Rev. D 40 (1989) 4178 - 4183.

\bibitem{Ivanov2001}
D. A. Ivanov, L. B.  Ioffe, V. B. Geshkenbein, and G. Blatter,
Interference effects in isolated Josephson junction arrays with geometric symmetries,
Phys. Rev. B 65 (2001) 024509.

\bibitem{Averin2002}
J. R. Friedman and D. V.  Averin,
Aharonov-Casher-Effect of Macroscopic Tunneling of Magnetic flux,
Phys. Rev. Lett. 88 (2002) 050403.

\bibitem{Matveev2002}
K. A. Matveev, A. I.  Larkin, and L. I. Glazman, 
Persistent Current in Superconducting Nanorings,
Phys. Rev. Lett. 89 (2002) 096802.

\bibitem{Pop2010} 
I. M. Pop, I. Protopopov, F. Lecocq, Z. Peng, B. Pannetier, O. Buisson and W. Guichard,
Measurement of the effect of quantum phase slips in a Josephson junction chain,
Nature Physics 6 (2010) 589--592.

\bibitem{Astafiev2012}
O. V. Astafiev, L. B. Ioffe, S. Kafanov, Yu. A. Pashkin, K. Yu. Arutyunov, 
D. Shahar, O. Cohen, and  J. S. Tsai,
Coherent quantum phase slip,
Nature Physics 484  (2012) 355--358.

\bibitem{Doniach1998}
A.\,Kr\"{a}mer and S.\,Doniach,
Superinsulator Phase of Two-Dimensional Superconductors,
\textit{Phys. Rev. Lett.} \textbf{81} 3523 - 3526 (1998).

\bibitem{Salzberg1963}
A.~Salzberg and S.~Prager, 
Equation of State for a Two-Dimensional Electrolyte, 
J. Chem. Phys. 38 (1963) 2587.

\bibitem{Berezinskii1970} 
V.~L.~Berezinskii, 
Violation of long range order in one-dimensional and two-dimensional 
systems with a continuous symmetry group. I. Classical systems, 
Zh. Eksp. Teor. Fiz. 59 (1970) 907--920 
[Sov. Phys. JETP 32 (1971) 493--500].

\bibitem{Berezinskii1971}
V.~L.~Berezinskii, 
Destruction of long-range order in one-dimensional and two-dimensional 
systems possessing a continuous symmetry group. II. Quantum systems, 
Zh. Eksp. Teor. Fiz. 61 (1971) 1144--1155 
[Sov. Phys. JETP 34 (1972) 610–-616].

\bibitem{KT1972}
J.~M.~Kosterlitz and D.~Thouless, 
Long range order and metastability in two dimensional solids and superfluids, 
J. Phys. C 5 (1972) L124--L126.
                                 
\bibitem{KT1973} 
J.~M.~Kosterlitz and D.~Thouless, 
Ordering, metastability and phase transitions in two-dimensional systems, 
J. Phys. C 6 (1973) 1181--1203.

\bibitem{Diamantini}
M.~C.~Diamantini, P.~Sodano, and C.~A.~Trugenberger, 
Gauge theories of Josephson junction arrays,  
Nuclear Physics B 474 (1996) 641--677.

\bibitem{Fisher1990}
M.~P.~A.~Fisher, G.~Grinstein, and S.~M.~Girvin,  
Presence of quantum diffusion in two dimensions: Universal resistance 
at the superconductor-insulator transition, 
Phys. Rev. Lett. 64 (1990) 587–-590.

\bibitem{Fisher1990B} 
M.~P.~A.~Fisher, 
Quantum phase transitions in disordered two-dimensional superconductors, 
Phys. Rev. Lett. 65 (1990) 923--926.


\bibitem{Demler2010} 
B.~I.~Halperin, G.~Refael, and E.~Demler,
Resistance in Superconductors, 
International Journal of Modern Physics B 24 (2010) 4039--4080.

\bibitem{ScrLengthCBKT}
Yukiya Miyachi and Susumu Kurihara, 
Effect of Finite Screening Length on Charge Kosterlitz-Thouless-Berezinskii Transition, 
J. Phys. Soc. Jpn. 69 (2000) 2356--2357. 

\bibitem{Khomskii1984} 
L.~N.~Bulaevskii, A.~A.~Sobyanin, and D.~I.~Khomskii, 
Superconducting properties of systems with local pairs, 
Zh. Eksp. Teor. Fiz. 87 (1984) 1490-1500 
[Sov. Phys. JETP 60 (1984) 856--862].

\bibitem{Khomskii1999}
M.~Yu.~Kagan, D.~I.~Khomskii, and M.~V.~Mostovoy, 
Double-exchange model: phase separation versus canted spins, 
Eur. Phys. J. B 12 (1999) 217--223.

\bibitem{Khomskii2001} 
M.~Yu.~Kagan, K.~I.~Kugel, and D.~I.~Khomskii, 
Phase separation in systems with charge ordering, 
JETP 93 (2001) 415--423.

\bibitem{Khomskii2008}
K.~I.~Kugel, A.~L.~Rakhmanov, A.~O.~Sboychakov, and D.~I.~Khomskii, 
Doped orbitally ordered systems: Another case of phase separation, 
Phys. Rev. B 78 (2008) 155113.

\bibitem{Khomskii2009}  
A.~O.~Sboychakov, K.~I.~Kugel, A.~L.~Rakhmanov,  and D.~I.~Khomskii, 
Phase separation in doped systems with spin-state transitions, 
Phys. Rev. B 80 (2009) 024423.

\bibitem{Dagotto} 
E.~Dagotto,
Complexity in Strongly Correlated Electronic Systems,
Science 309 (2005) 257--262.

\bibitem{Islands} 
A.~Glatz, I.~S.~Aranzon, T.~I.~Baturina, N.~M.~Chtchelkatchev, and  V.~M.~Vinokur, 
Self-organized superconducting textures in thin films, 
Phys. Rev. B 84 (2011) 024508.

\bibitem{Syzranov2010} 
S.~V.~Syzranov, I.~L.~Aleiner, B.~L.~Altshuler, and K.~B.~Efetov, 
Coulomb Interaction and First-Order Superconductor-Insulator Transition, 
Phys. Rev. Lett. 105 (2010) 137001.

\bibitem{Rytova1967}
N.~S.~Rytova,
Screening potential of the point charge in a thin film,
Vestnik MSU 3 (1967) 30--37 (in Russian).

\bibitem{ChaplikEntin1971}
A.~V.~Chaplik, M.~V.~Entin,
Charged impurities in very thin layers,
Zh. Eksp. Teor. Fiz. 61 (1971) 2496--2503.

\bibitem{Keldysh1979}
L.~V.~Keldysh, 
Coulomb interaction in thin semiconductor and semimetal films,
JETP Lett. 29 (1979) 658--661. 

\bibitem{Dubrov} 
V.\,E.~Dubrov, M.\,E.~Levinstein, and  M.\,S.~Shur, 
Permittivity anomaly in metal-dielectric transitions. Theory and simulation,  
Zh. Eksp. Teor. Fiz. 70 (1976) 2014--2024
[Sov. Phys. JETP 43 (1976) 1050--1056].

\bibitem{Stauffer} 
D.~Stauffer and A.~Aharony,
Introduction to Percolation Theory,
2nd ed., Taylor and Francis, London 1994 (second printing).

\bibitem{Castner1975}
 T.\,G.\,Castner, N.\,K.\,Lee, G.\,S.\,Cieloszyk, and G.\,L.\,Salinger,
 Dielectric Anomaly and the Metal-Insulator Transition in n-Type Silicon,
 Phys. Rev. Lett. 34 (1975) 1627-1630.

\bibitem{Herzfeld1927}
K.\,F.\,Herzfeld,
On atomic properties which make an element a metal,
Phys. Rev. 29 (1927) 701-705.

\bibitem{Shimshoni1998} 
E.~Shimshoni, A.~Auerbach, and A.~Kapitulnik, 
Transport through Quantum Melts, 
Phys. Rev. Lett. 80 (1998) 3352--3355.

\bibitem{Rosenbaum} 
H.~F.~Hess, K.~DeConde, T.~F.~Rosenbaum,  and G.~A.~Thomas, 
Giant dielectric constants at the approach to the insulator-metal transition, 
Phys. Rev. B 25 (1982) 5578--5580.

\bibitem{Shalnikov38} 
A.~I.~Shal'nikov, 
Superconducting Thin Films, 
Nature (London) 142 (1938) 74.

\bibitem{Shalnikov40} 
A.~I.~Shal'nikov, 
Superconducting Properties of Thin Metallic Layers, 
Zh. Eksp. Teor. Fiz. 10 (1940) 630--640. 

\bibitem{Dynes1986} 
R.~C.~Dynes, A.~E.~White, J.~M.~Graybeal, and J.~P.~Garno, 
Breakdown of Eliashberg Theory for Two-Dimensional Superconductivity in the Presence of Disorder, 
Phys. Rev. Lett. 57 (1986) 2195--2198.

\bibitem{Xiong1995}
P.~Xiong, A.~V.~Herzog, and R.~C.~Dynes, 
Superconductivity in ultrathin quench-condensed Pb/Sb and Pb/Ge multilayers, 
Phys. Rev. B 52 (1995) 3795--3801.  

\bibitem{Raffy83} 
H.~Raffy, R.~B.~Laibowitz, P.~Chaudhari, and S.~Maekawa, 
Localization and interaction effects in two-dimensional W-Re films, 
Phys. Rev. B 28 (1983) 6607--6609.

\bibitem{Graybeal84} 
J.~M.~Graybeal and M.~R.~Beasley, 
Localization and interaction effects in ultrathin amorphous superconducting films, 
Phys. Rev. B 29 (1984) 4167--4169. 

\bibitem{Rogachev2012}
Hyunjeong Kim, Anil Ghimire, Shirin Jamali, Thaddee K. Djidjou, Jordan M. Gerton, and A. Rogachev, 
Effect of magnetic Gd impurities on the superconducting state of amorphous Mo-Ge thin films
with different thickness and morphology,
Phys. Rev. B 86 (2012) 024518. 

\bibitem{Hebard1985}
A.~F.~Hebard and M.~A.~Paalanen, 
Diverging Characteristic Lengths at Critical Disorder in Thin-Film Superconductors, 
Phys. Rev. Lett. 54 (1985) 2155--2158.

\bibitem{Okuma1998} 
S.~Okuma, T.~Terashima, and N.~Kokubo, 
Anomalous magnetoresistance near the superconductor-insulator transition
in ultrathin films of $a$-Mo$_x$Si$_{1-x}$, 
Phys. Rev. B 58 (1998) 2816--2819. 

\bibitem{Yoon2006} 
Y.~Qin, C.~L.~Vicente, and J.~Yoon,
Magnetically induced metallic phase in superconducting tantalum films, 
Phys. Rev. B 73 (1998) 100505(R). 

\bibitem{Aubin2008} 
C.~A.~Marrache-Kikuchi, H.~Aubin, A.~Pourret, K.~Behnia,  J.~Lesueur, 
L.~Berg\'{e}, and L.~Dumoulin, 
Thickness-tuned superconductor-insulator transitions under magnetic field in $a$-NbSi, 
Phys. Rev. B 78 (2008) 144520. 

\bibitem{Baturina2011} 
T.~I.~Baturina, S.~V.~Postolova, A.~Yu.~Mironov, A.~Glatz, M.~R.~Baklanov, and V.~M.~Vinokur, 
Superconducting phase transitions in ultrathin TiN films,  
EPL 97 (2012) 17012.

\bibitem{TiNJapan2000} 
T.~Suzuki, Y.~Seguchi, and T.~Tsuboi, 
Fermi Liquid Effect on Tricritical Superconducting Transitions
in Thin TiN Films under the Spin Paramagnetic Limitation,
J. Phys. Soc. Jpn. 69 (2000) 1462--1471.

\bibitem{OtoPtSi1994} 
K.~Oto, S.~Takaoka, and K.~Murase, 
Superconductivity in PtSi ultrathin films, 
J. Appl. Phys. 76 (1994) 5339--5342.

\bibitem{Maekawa} 
S.~Maekawa and H.~Fukuyama, 
Localization effects in two-dimensional superconductors, 
J. Phys. Soc. Jpn. 51 (1982) 1380--1385.

\bibitem{Finkelstein} 
A.~M. Finkel'stein, 
Superconducting transition temperature in amorphous films, 
Pis'ma Zh. Eksp. Teor. Fiz. 45 (1987) 37--40
[Sov. Phys. JETP Lett. 45 (1987) 46--49];
Suppression of superconductivity in homogeneously disordered systems, 
Physica B 197 (1994) 636--648.

\bibitem{HebardKotliar1989}
A.~F.~Hebard and G.~Kotliar, 
Possibility of the vortex-antivortex transition temperature of a thin-film superconductor being renormalized by disorder,
Phys. Rev. B 39 (1989) 4105 -- 4109. 

\bibitem{Mott1937}
N.~F.~Mott and R.~Peierls, 
Discussion of the paper by de Boer and Verwey,
Proceedings of the Physical Society of London 49 (1937) 72.

\bibitem{Mott1949}
N.~F.~Mott,
The basis of the electron theory of metals, with special reference to the transition metals,
Proceedings of the Physical Society of London, Ser. A 62 (1949) 416.

\bibitem{Anderson1958}
P.~W.~Anderson,
Absence of Diffusion in Certain Random Lattices,
Phys. Rev. 109 (1958) 1492--1505.

\bibitem{Kohn1964} 
W.~Kohn, 
Theory of Insulating State,  
Phys. Rev. 133 (1964) A171--A181 .

\bibitem{Abrahams1979}
E.~Abrahams, P.~W.~Anderson, D.~C.~Licciardello, T.~V.~Ramakrishnan,
Scaling Theory of Localization: Absence of Quantum Diffusion in Two Dimensions,
Phys. Rev. Lett. 42 (1979) 673–-676.

\bibitem{ImadaReview} 
Masatoshi Imada,  Atsushi Fujimori, and  Yoshinori Tokura,
Metal-insulator transitions,  
Rev. Mod. Phys. 70 (1998) 1039–-1263.

\bibitem{Brandes2003}
T.~Brandes and  S.~Kettemann, 
The Anderson Transition and its Ramifications --- Localisation, 
Quantum Interference, and Interactions, 
Berlin: Springer Verlag, 2003.

\bibitem{VinLark} 
A.~I.~Larkin and V.~M.~Vinokur, 
Bose and Vortex Glasses in High Temperature Superconductors, 
Phys. Rev. Lett. 75 (1995) 4666--4669.

\bibitem{Basko2006} 
D.~M.~Basko, I.~L.~Aleiner, and B.~L.~Altshuler, 
Metal-insulator transition in a weakly interacting many-electron system with
localized single-particle states, 
Ann. Phys. 321 (2006) 1126--1205.

\bibitem{Gornyi} 
I.~V.~Gornyi, A.~D.~Mirlin, and D.~G.~Polyakov, 
Interacting Electrons in Disordered Wires: Anderson Localization and Low-$T$ Transport,
Phys. Rev. Lett. 95 (2005) 206603.

\bibitem{BLV} 
I.~S.~Beloborodov, A.~V.~Lopatin, and V.~M.~Vinokur, 
Coulomb effects and hopping transport in granular metals, 
Phys. Rev. B 72  (2005) 125121.

\bibitem{Falko2009} 
G.~M.~Falco, T.~Nattermann, and V.~L.~Pokrovsky, 
Weakly interacting Bose gas in a random environment, 
Phys. Rev. B 80 (2009) 104515.

\bibitem{AAS2010} 
I.~L.~Aleiner, B.~L.~Altshuler, and G.~V.~Shlyapnikov, 
A finite-temperature phase transition for disordered weakly interacting bosons in
one dimension, 
Nature Phys. 6 (2010) 900--904. 

\bibitem{Lopatin2007} 
A.~V.~Lopatin and V.~M.~Vinokur,
Hopping transport in granular superconductors,
Phys. Rev. B 75 (2007) 092201.

\bibitem{CVB2009}  
N.~M.~Chtchelkatchev,  V.~M.~Vinokur, and T.~I.~Baturina, 
Hierarchical Energy Relaxation in Mesoscopic Tunnel Junctions:
Effect of a Nonequilibrium Environment on Low-Temperature Transport, 
Phys. Rev. Lett. 103 (2009) 247003.

\bibitem{CVBPhysC2010} 
N.~M.~Chtchelkatchev,  V.~M.~Vinokur, and T.~I.~Baturina, 
Nonequilibrium transport in superconducting tunneling structures, 
Physica C 470 (2010) S935-S936.

\bibitem{CVB2011NATO}  
N.\,M.~Chtchelkatchev,  V.\,M.~Vinokur, and T.\,I.~Baturina, 
Low temperature transport in tunnel junction arrays: Cascade energy relaxation, 
in: Physical Properties of Nanosystems, edited by J.~Bonca and S.~Kruchinin,
NATO Science for Peace and Security Series B: Physics and Biophysics
(Springer Science+Business Media B.V., Dordrecht, 2011), chapter 3, 
25--44.

\bibitem{Ioffe2010} 
L.~B.~Ioffe and M.~M\'{e}zard, 
Disorder-Driven Quantum Phase Transitions in Superconductors and Magnets,
Phys. Rev. Lett. 105 (2010) 037001.

\bibitem{FIM2010} 
M.~V.~Feigel'man,  L.~B.~Ioffe, and M.~M\'{e}zard, 
Superconductor-insulator transition and energy localization, 
Phys. Rev. B 82 (2010) 184534.

\bibitem{Ingold1991} 
G.-L.~Ingold and Yu.~V.~Nazarov, 
Charge Tunneling Rates in Ultrasmall Junctions, 
in Single Charge Tunneling, edited by H.~Grabert 
and M.~H.~Devoret, NATO ASI, Ser. B, Vol. 294 (Plenum, New York,
1991). Chapter 2, 21--107.

\bibitem{Tinkham} 
M.~Tinkham, D.~W.~Abraham, and C.~J.~Lobb, 
Periodic flux dependence of the resistive transition in two-dimensional superconducting arrays, 
Phys. Rev. B 28 (1983) 6578--6581. 

\bibitem{ESITnmat2012}
Adrien Allain, Zheng Han, and Vincent Bouchiat,
Electrical control of the superconducting-to-insulating transition in graphene-metal hybrids,
Nature Materials 11 (2012) 590--594.

\bibitem{ReviewElectrostatics}
C. H. Ahn, A. Bhattacharya, M. Di Ventra, J. N. Eckstein, C. Daniel Frisbie, M. E. Gershenson, 
A. M. Goldman, I. H. Inoue, J. Mannhart, Andrew J. Millis, Alberto F. Morpurgo, Douglas Natelson, 
Jean-Marc Triscone,
Electrostatic modification of novel materials,
Rev. Mod. Phys. 78 (2006) 1185--1212.

\bibitem{GoldmanESIT}
Kevin A. Parendo, K. H. Sarwa B. Tan, and A. M. Goldman,
Electrostatic and parallel-magnetic-field tuned two-dimensional
superconductor-insulator transitions,
Phys. Rev. B 73 (2006) 174527. 

\bibitem{Matthey2007}
D. Matthey, N. Reyren, J.-M. Triscone, and T. Schneider,
Electric-field-effect modulation of the transition temperature, 
mobile carrier density, and in-plane penetration depth of NdBa$_2$Cu$_3$O$_{7-\delta}$ thin films, 
Phys. Rev. Lett. 98 (2007) 057002.

\bibitem{Caviglia2008}
A. D. Caviglia, S. Gariglio, N. Reyren, D. Jaccard, T. Schneider, M. Gabay, S. Thiel, G. Hammerl, 
J. Mannhart, and J.-M. Triscone, 
Electric field control of the LaAlO$_3$/SrTiO$_3$ interface ground state, 
Nature 456 (2008) 624--627.  

\bibitem{Ueno2008} 
K. Ueno, S. Nakamura, H. Shimotani, A. Ohtomo, N. Kimura, T. Nojima, H. Aoki, 
Y. Iwasa, and M. Kawasaki,
Electric-field-induced superconductivity in an insulator, 
Nature Materials 7 (2008) 855--858.

\bibitem{Bell2009}
C. Bell, S. Harashima, Y. Kozuka, M. Kim, B. G. Kim, Y. Hikita, and H. Y. Hwang,
Dominant Mobility Modulation by the Electric Field Effect at the LaAlO$_3$/SrTiO$_3$ Interface,
Phys. Rev. Lett. 103 (2009) 226802.

\bibitem{Biscaras2010}
J. Biscaras,	 N. Bergeal, A. Kushwaha, T. Wolf, A. Rastogi, R. C. Budhani, and  J. Lesueur,
Two-dimensional superconductivity at a Mott insulator/band insulator 
interface LaTiO$_3$/SrTiO$_3$,
Nature Commun. 1 (2010) 89.

\bibitem{Ye2010}
J. T. Ye, S. Inoue, K. Kobayashi, Y. Kasahara, H. T. Yuan, H. Shimotani, and Y. Iwasa,
Liquid-gated interface superconductivity on an atomically flat film,
Nature Materials 9 (2010) 125--128.

\bibitem{Dhoot2010}
Anoop Singh Dhoot, Stuart C. Wimbush, Tim Benseman, Judith L. MacManus-Driscoll, 
J. R. Cooper, and Richard Henry Friend, 
Increased Tc in electrolyte-gated cuprates. 
Adv. Mater. 22 (2010) 2529--2533.

\end{thebibliography}
\end{document}